\documentclass[authoryear]{article}
\usepackage[utf8]{inputenc}
\usepackage{hyperref}
\usepackage{cancel}
\usepackage{graphicx}
\usepackage{xcolor}
\usepackage{ulem}
\usepackage{amsmath,amssymb}
\usepackage{mathrsfs}
\usepackage{bm}
\usepackage{lineno}
\usepackage{siunitx}
\usepackage{booktabs}

\usepackage{natbib}
\setcitestyle{authoryear,open={(},close={)}}

\usepackage{amsfonts} 

\DeclareMathSymbol{\shortminus}{\mathbin}{AMSa}{"39}

\usepackage{myglossaries}
\glsdisablehyper
\usepackage[left=3cm,bottom=4cm,right=3cm,top=4cm]{geometry}

\title{A single oblate spheroid settling in unbounded ambient fluid:  \\
       a benchmark for simulations in steady \\
       and unsteady wake regimes}

\date{{\it Accepted in IJMF 2020}}

\def\affilKar{Institute for Hydromechanics, Karlsruhe Institute of Technology,
              Germany}
\def\affilJan{Universit\'e de Strasbourg, Fluid Mechanics Department, 
              Institut ICube, France}

\author{Manuel Moriche\thanks{\affilKar}  \\
        Markus Uhlmann\footnotemark[1] \\
        Jan Du\v{s}ek\thanks{\affilJan}}

\usepackage{ulem}
\newcommand\review[2]{#2}

\newcommand{\ddt}[1]{\frac{\text{d}#1}{\text{d}t}}
\newcommand{\ppt}[1]{\frac{\partial #1}{\partial t}}

\newcommand{\wrt}{\text{d}}

\begin{document}

\maketitle
\begin{abstract}
We have performed spectral/spectral-element simulations of a single
oblate spheroid with small geometrical aspect ratio settling in an
unbounded ambient fluid, for a range of Galileo numbers covering the
various regimes of motion (steady vertical, steady oblique,
\review{vertically oscillating}{vertical periodic} and chaotic).
The high-fidelity data provided includes particle quantities
(statistics in the chaotic case), as well as flow profiles and
pressure maps. 
The reference data can be used as
\review{a}{an additional} 
benchmark 
for other numerical approaches, where a careful grid convergence 
study for a specific target parameter point is often useful.
%
\review{We further describe an extension of the immersed boundary method of
  Uhlmann (J. Comput. Phys, 209(2):448--476, 2005) to enable the
  simulation of non-spherical particles.}{%
  We further describe an extension of a specific immersed boundary
  method (Uhlmann, J. Comput. Phys, 209(2):448--476, 2005) to enable
  the tracking of non-spherical particles.}
%
Finally, the reference cases are computed with this immersed boundary
method at various spatial and temporal resolutions, and grid
convergence is discussed over the various regimes of spheroidal
particle motion.
The cross-validation results can serve as a guideline for the design
of simulations with the aid of similar non-conforming methods, 
involving spheroidal particles with Galileo numbers of
${\cal O}(100)$. 
\end{abstract}

\section{Introduction}
\label{sec:intro}

The flow around a single particle settling/rising in an otherwise quiescent 
fluid has received the attention from the scientific community for many years 
\citep{ern:2012}.
The non-trivial particle paths which result in these flows play a significant 
role in a variety of fields such as meteorology, sedimentology or bio-inspired 
aerodynamics.
The complex dynamics that govern particle-fluid interactions are influenced by
fluid properties, particle shape, inertia, and gravity.
As a consequence, the size of the parameter space of the problem is quite 
large, and, despite the progress made in the last decades, it has still not been
fully explored in all detail.
When the Reynolds number of the flow around the particle exceeds values of 
${\cal O}(10)$, wake effects become significant.
In this scenario, analytical solutions are not available, and in engineering
problems one needs to resort to empirical correlations in order to predict the
hydrodynamic forces acting on particles \citep{balachandar:2010}.
Therefore, experiments and/or numerical simulations in which the flow and the 
particle motion are fully resolved are one way to drive further model 
development.

When a single sphere is fixed in a uniform flow, different flow regimes appear 
as a function of the Reynolds number based on the free stream velocity and the
sphere's diameter \citep{natarajan:1993,tomboulides:2000}.
For freely-mobile particles the Reynolds number is not a predetermined parameter
any more, but the result of the balance of hydrodynamic and gravitational 
forces.
In this situation, the resultant flow regime is uniquely defined by the density
ratio of the two phases, $\gls{kappa}=\gls{rhop}/\gls{rhof}$, and the Galileo 
number $\gls{Ga}=\gls{Ug}\gls{dd}/\gls{nu}$ (where $\gls{Ug}=\sqrt{\left(%
\gls{kappa}-1\right)\gls{g}\gls{dd}}$ is a characteristic gravitational 
velocity, \gls{g} is the gravitational acceleration, \gls{dd} the sphere's 
diameter, and \gls{nu} the kinematic viscosity of the fluid).
Spheres have been widely studied in the literature 
\citep{jenny:2004,veldhuis:2007,uhlmann:2014b} mainly because the simplicity of
their shape allows to reduce the complexity of the problem.
However, particles found in real flows are often non-spherical.

One possible single-parameter deviation from a spherical shape is a spheroid,
also called ellipsoid of revolution.
A spheroid is non-isotropic, but axi-symmetric, and is uniquely defined by its 
equatorial diameter \gls{dd} and its aspect ratio $\gls{chi}=\gls{dd}/\gls{aa}$, 
where \gls{aa} is the length of the symmetry axis of the spheroid.
In a very simple way one can obtain flat, disk-like particles by setting 
$\gls{chi}>1$ (oblate spheroid), elongated, rod-like particles with 
$\gls{chi}<1$ (prolate spheroid) or recover a sphere with $\gls{chi}=1$.
This implies that the flow regimes appearing when a fixed spheroid is placed in
a free stream are defined by the Reynolds number \gls{Re} and \gls{chi} 
\citep{chrust:2010}; when the spheroid is released the problem is defined by 
the triplet \gls{Ga}, \gls{kappa} and \gls{chi}.
The case of fixed oblate spheroids has been investigated by \cite{chrust:2010} 
who provided a detailed analysis of the flow transition as a function of the 
geometrical aspect ratio.
The dynamics of free falling oblate spheroids have recently been analyzed by 
\cite{zhou:2017}.
They found that the two parameter $\gls{Ga}$-$\gls{kappa}$ state diagram for a
given aspect ratio~\gls{chi} features a higher degree of complexity than that of a
sphere.
Furthermore, they found strong quantitative and qualitative differences for 
different \gls{chi}.
Overall, the state diagram for $\gls{chi}=1.1$ resembles some features of 
spheres and the state diagram for $\gls{chi}=10$ resembles that of a flat disk
\citep{chrust:2013}.
For intermediate values, there are strong differences between $\gls{chi}=1.1$
and $\gls{chi}=2$, whereas the transition in the behavior from $\gls{chi}=2$ to
$10$ is somehow progressive.
We will discuss these various flow states in more detail in 
\S~\ref{sec:overview} below.

Understanding the underlying physics of the settling/rising of a single 
suspended particle also serves as a basis for understanding collective effects
which arise in corresponding multi-particle systems.
For example, \cite{uhlmann:2014a} related the onset of clustering of a dilute 
set of settling spheres in unbounded fluid with the transition of wake regimes
in the case of an isolated sphere.

The cost of simulating the flow around freely moving particles by means of 
particle resolved \gls{dns} is very high, but thanks to the efficiency of 
non-conforming algorithms like the \gls{ibm} \citep[][]{mittal:2005} or the 
Lattice-Boltzmann method \citep{chen:1998}, simulations of the flow around~${\cal O}(10^{6})$
particles can be found in the literature \citep{kidanemariam:2017}.
However, these algorithms based on non-conforming meshes require careful 
validation and grid-convergence analysis, which is a difficult
undertaking
\review{when significant relative velocities between the solid and the
fluid phase are to be simulated.}{%
when systems featuring significant particle wakes are to be
simulated.} 
%
\review{A discussion of the lack of representative features found in typical
validation processes for the case of spherical particles}{%
A discussion of how the validation process of
the flow around
spheres could benefit from
detailed reference data at higher particle Reynolds numbers} 
can be found in \cite{uhlmann:2014b}.
\review{The case of non-spherical particles is in no better position.
The most common approach found in the literature is to use the analytical 
solution of the motion of an ellipsoid in shear flow for $\gls{Re}\ll
1$ obtained by  Jeffery, 1992
 (Aidun et al., 1998; Huang et al. 2012; Eshghinejadfard et al., 2016;
Ardekani et al., 2016).
Therefore, we believe that a thorough validation process in which
non-conforming grid algorithms are validated in order to realize all
their potential in large multi-particle problems with non-spherical
shapes  is missing in the literature. In this context the
highly-accurate \gls{sem} proposed by  Chrust et al. (2013)  to simulate
the flow around settling spheroids meets the requirements of accuracy
and capability of representing relevant flow conditions, and it will
therefore be employed herein.}{%
Regarding particles with spheroidal shape, 
many numerical studies found in the literature resort to the analytical 
solution describing the rotational motion of a particle placed in
shear flow under creeping flow conditions \citep[obtained
by][]{jeffery:1922} for the purpose of validation
\citep{aidun:1998,huang:2012,eshghinejadfard:2016,ardekani:2016}.
On the other hand, the studies of \cite{chrust:2013} and 
\cite{zhou:2017} have demonstrated that high-fidelity data for
settling rigid bodies with an axi-symmetric shape
can be  generated with the aid of a body-conforming \gls{sem}. 
Previously, \cite{tschisgale:2018} have performed a validation study
of a (non-grid-conforming) immersed boundary method with the aid
of spectral-element data for the case of a light oblate spheroid and
for a heavy thin disc,
while \cite{arranz:2018} used spectral-element data for a settling
oblate spheroid for the same purpose. 
In the present work we provide additional benchmark data for settling
spheroids at several parameter points, including detailed information
on the particle motion as well as on the hydrodynamic fields. Our goal
is to complement the existing set of reference data in order to
further facilitate the development of efficient numerical methods.}

\review{%
  In this work we present spectral-element solutions of spheroids with a small 
  aspect ratio ($\gls{chi}=1.1$, almost spherical shape) and of
  spheroids with a moderate aspect ratio ($\gls{chi}=1.5$). 
  These two aspect ratios represent spheroids whose \gls{Ga}-\gls{kappa} state
  diagram still maintains similarities with that of the sphere prior to the 
  qualitative difference found for $\gls{chi}\geq 2$.
  This allows future studies on the settling of non-spherical
  particles to relate their results with the widely studied problem of
  settling spheres. 
  Furthermore, we describe an extension of the \gls{ibm} proposed by 
  Uhlmann (2005) to simulate the flow around non-spherical
  particles.}{%
  The present work focuses on settling spheroids with
  two different geometrical aspect ratios: 
  $\gls{chi}=1.1$ with an almost spherical shape, and 
  a moderate value $\gls{chi}=1.5$. 
  For these two geometries the respective \gls{Ga}-\gls{kappa} state
  diagrams still maintain similarities with the one of the sphere,
  since qualitative differences are observed for $\gls{chi}\geq 2$. 
  Beyond the immediate use as validation data, this choice allows future
  studies on the settling of non-spherical particles to relate their
  results with the widely studied problem of settling spheres. 

  Furthermore, in the present work we discuss an extension of one
  specific immersed-boundary method for the simulation of particulate flow
  \citep{uhlmann:2005} such that the motion of non-spherical 
  particles can be tackled. The modifications with respect to the
  orginal algorithm concern two aspects: the tracking of rotational
  motion (involving a quaternion description, as well as its temporal
  discretization), and the distribution of Lagrangian force poins per
  particle.
  It should be noted that a number of authors have previously proposed 
  immersed-boundary methods valid for the flow around mobile,
  non-spherical particles
  \citep[e.g.][and
  others]{yang:15,eshghinejadfard:2016,ardekani:2016,tschisgale:2018}.
  Generally speaking, the present method is kept as simple as possible,  
  and algorithmic differences compared to alternative techniques are
  provided where appropriate in the technical description below.  
}

\review{%
  The purpose of this work can be divided into a) provide high-fidelity data of 
  settling spheroids of low and moderate aspect ratio with significant relative 
  velocities between the fluid and the particle, b) introduce an algorithm to 
  handle non-spherical particles in an existing \gls{ibm} algorithm originally 
  proposed for spherical particles, and c) present an extensive validation process
  to the mentioned algorithm that can be used in the context of any numerical 
  approach whose target problem is similar to the one analyzed here.}{%
  The overall purpose of the present work is, therefore, threefold:
  (a) provide high-fidelity data of settling spheroids of low and
  moderate aspect ratio at particle Reynolds numbers ${\cal O}(100)$; 
  (b) extend the existing \gls{ibm} algorithm of Uhlmann (2005) such
  that non-spherical particle motion can be handled efficiently; 
  (c) present an extensive validation process that can in principle be
  used in the context of a wide range of numerical methods for the 
  simulation of submerged, mobile, rigid bodies. 
}

The manuscript is organized as follows.
In \S~\ref{sec:problem} the setup of the problem under
\review{study}{investigation} 
is presented
together with an overview of the flow regimes that a free-falling spheroid can 
encounter.
In \S~\ref{sec:sem} the
\review{spectral-element}{spectral/spectral-element} 
approach is described together with the
reference solutions obtained.
In \S~\ref{sec:ibm/method} the extension of the immersed boundary method of 
\cite{uhlmann:2005} is described and validated in \S~\ref{sec:ibm} against the 
spectral-element reference data.

\section{Problem description}                               
\label{sec:problem}

We consider the settling of a single particle under the action of gravity in an
incompressible Newtonian fluid.
The particle is a rigid oblate spheroid with equatorial diameter \gls{dd}, 
aspect ratio $\gls{chi}=\gls{dd}/\gls{aa}$, where \gls{aa} is the length of the 
symmetry axis, and uniform density \gls{rhop}.
Two Cartesian coordinate systems are used in this work.
The first one is an inertial reference frame \gls{Ofix}, in which the fluid
velocity is zero in the absence of the particle.
The vertical axis of this coordinate system $z_\text{fix}$ is represented by the
unit vector $\gls{ez}=-\gls{vg}/\gls{g}$, where $\gls{g}=\left|\gls{vg}\right|$
is the magnitude of the gravitational acceleration vector \gls{vg}.
Second, a non-inertial reference frame \gls{Oxyz} is aligned with \gls{Ofix},
and it has an origin which is attached to the particle's center of gravity.

\begin{figure}
\begin{center}
\includegraphics{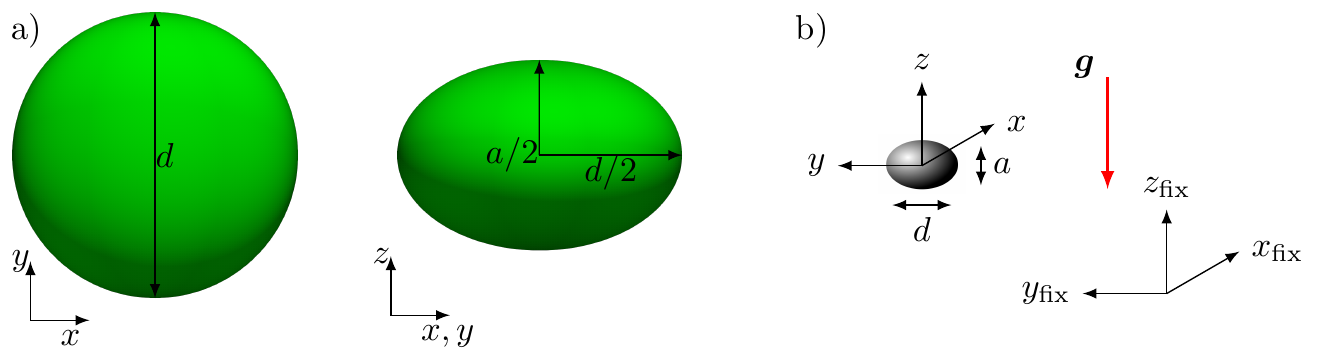}
\caption{a) View along the symmetry axis and perpendicular to it of the spheroid
with equatorial diameter \gls{dd} and length of symmetry axis \gls{aa} and b) 
sketch of the spheroid in the unbounded domain.
\label{fig:problem_description}}
\end{center}
\end{figure}

The governing equations for the flow field are the Navier-Stokes equations for
an incompressible fluid
\begin{subequations}\label{eq:gov}
\begin{align}
\ppt{\gls{vu}}+\left(\gls{vu}\cdot\nabla\right)\gls{vu}
              +\nabla \gls{p}&= \gls{nu} \nabla^2\gls{vu},
                \label{eq:gov_mom} \\
\nabla\cdot\gls{vu}&=0 \label{eq:gov_cont},
\end{align}
\end{subequations}
where \gls{p} is the \review{hydrodynamic pressure}{pressure normalized by the
fluid density}, $\gls{vu}=\left(\gls{vu_x},%
\gls{vu_y},\gls{vu_z}\right)$ is the fluid velocity expressed in the inertial
reference frame \gls{Ofix} and \gls{nu} is the fluid's kinematic viscosity.
The no slip \review{}{and no penetration} condition is imposed at the boundary
of the particle.
The position and velocity of the particle are obtained by integration of the 
Newton-Euler equations of rigid body motion \review{}{over the surface of the
particle $S$}
%
\begin{subequations}\label{eq:motion}
\begin{align}
\gls{vol}\gls{rhop}\ddt{\gls{vup}}&=\gls{rhof}\int_{S} \gls{tau}\cdot\vec{n} \, \wrt \sigma
                            +\left(\gls{rhop}-\gls{rhof}\right)\gls{vol}\gls{vg}, \label{eq:motion_lin}\\
\ddt{\left(\gls{I}\gls{vomep}\right)}&= %
 \gls{rhof}\int_{S} \gls{rs}\times \left(\gls{tau}\cdot\vec{n}\right) \wrt \sigma, \label{eq:motion_rot}
\end{align}
\end{subequations}
%
where $\gls{vup}=\left(\gls{vup_x},\gls{vup_y}, \gls{vup_z}\right)$ and 
$\gls{vomep}=\left(\gls{vomep_x},\gls{vomep_y}, \gls{vomep_z}\right)$ are the 
linear and angular velocity of the particle expressed in the inertial reference
frame \gls{Ofix}, respectively, 
\review{$\gls{tau} = p\,\tens{I} + \gls{nu}\left(\nabla\gls{vu}+\nabla\gls{vu}^T\right)$ 
is the viscous stress}{%
$\gls{tau} =-p\,\tens{I} + \gls{nu}\left(\nabla\gls{vu}+\nabla\gls{vu}^T\right)$ 
is the hydrodynamic stress
tensor ($\tens{I}$ is the identity matrix), $\gls{vol}=\pi%
\gls{dd}^3/\left(6\gls{chi}\right)$ is the volume of the particle, 
\gls{I} is the inertia tensor,} $\vec{n}$ is a unit vector normal to the surface of 
the particle pointing towards the fluid, and \gls{rs} is the position vector of
a point on the surface of the particle with respect to its center.

Dimensional analysis shows that the problem is governed by three non-dimensional
parameters, namely the aspect ratio \gls{chi}, the density ratio between the
particle and the fluid $\gls{kappa}=\gls{rhop}/\gls{rhof}$ and the Galileo 
number $\gls{Ga}=\gls{Ug}\gls{dd}/\gls{nu}$, where \gls{Ug} is \review{the gravitational
velocity}{a gravitational velocity scale}
defined for heavy particles ($\gls{kappa}>1$) as $\gls{Ug}=\sqrt{\left(
\gls{kappa}-1\right)\gls{g}\gls{vol}/\gls{dd}^2}$.
The definition of \gls{Ug} and hence \gls{Ga} is the same as in 
\cite{chrust:2013} and \cite{zhou:2017}.
In the case of spheres, previous authors have typically rather used $\gls{Ga2}=%
\gls{Ug2}\gls{dd}/\gls{nu}$, where $\gls{Ug2}=\sqrt{\left(\gls{kappa}-1\right)%
\gls{g}\gls{dd}}$, which differ by factors $\sqrt{\pi/6}$ and $\gls{chi}^{2/3}
\sqrt{\pi/6}$, respectively, from the above counterparts.
We also introduce the non-dimensional mass $\gls{mstar}=\gls{kappa}\pi/\left(
6\gls{chi}\right)$, which has been used in previous work in the literature since it
provides a somewhat more straightforward scaling of the path regime boundaries
than the density ratio itself.
Note that as shown by \cite{chrust:2013} the steady and non-rotating regimes 
(steady vertical and steady oblique) are independent of the non-dimensionalized 
mass or, equivalently, the density ratio.

\subsection{Overview of flow regimes and dynamics of oblate spheroids}
\label{sec:overview}

In this section we present a brief description of the flow regimes that appear
when an oblate spheroid settles in an unbounded fluid under the action of
gravity.
This presents a summary of the relevant results of \cite{zhou:2017}.

For each aspect ratio \gls{chi} the main characteristics of the two parameter 
(\gls{Ga}, \gls{kappa}) state diagram are presented, focusing on the appearance
of the sphere-like scenario, flutter and tumbling modes.
The sphere-like scenario corresponds to a successive transition from the steady
axisymmetric regime to a steady oblique regime and, eventually, to unsteady 
regimes.
This transition scenario is found for all density ratios in the case of spheres
\citep{jenny:2004}.
Flutter is characterized by a periodic swinging motion with significant 
amplitude but with deviations from the vertical of less than $90$ degrees.
Tumbling is a rotating motion during which the angular velocity varies 
periodically without changing sign.

\begin{figure}
\begin{center}
\includegraphics[scale=1.0]{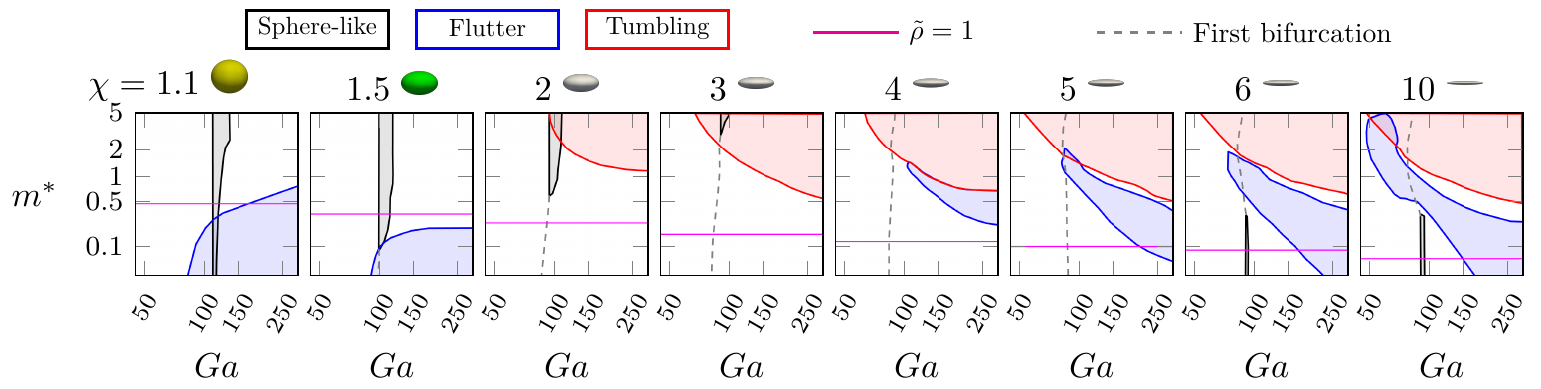}
\caption{Outline of the state diagrams for different aspect ratios. The 
horizontal magenta line corresponds to $\gls{kappa}=1$, and the coloured regions
indicate the occurrence of sphere-like, flutter or tumbling motion (see legend). 
The first bifurcation from the steady vertical regime is also indicated by a 
dashed gray line. The represented sphere-like domain includes the steady oblique
and oblique oscillating regimes. The range of values shown in all the panels is 
$45<\gls{Ga}<300$ and $0<\gls{mstar}<5$.
$\gls{Ga}$ is represented in logarithmic scale whereas the represented values of
\gls{mstar} are linear in $\log{\gls{mstar}+0.1}$. 
The complete diagrams for all aspect ratios except $\gls{chi}=1.5$ can be found
in \cite{zhou:2017}. For a more detailed view of the case with $\gls{chi}=1.5$
see the present figure \ref{fig:regimesdetail}.
\label{fig:regimes_allchi}}
\end{center}
\end{figure}

Figure \ref{fig:regimes_allchi} shows an outline of the main features of the 
(\gls{Ga}, \gls{kappa}) state diagrams discussed here.
For spheroids with aspect ratio as low as $\gls{chi}=1.1$ the sphere-like 
scenario is present for all values of $\gls{kappa}>1$.
Flutter is observed for light particles and for small excess particle 
densities if $\gls{Ga}\gtrsim 200$, while tumbling is not observed.
For $\gls{chi}=1.5$ the sphere-like scenario is still observed for heavy 
particles and for light particles if $\gls{kappa} \gtrsim 0.3$.
The region in which flutter is observed is shrunk compared to $\gls{chi}=1.1$,
and there is no tumbling.
When the aspect ratio is increased to $\gls{chi}=2$ and $3$, the lower bound of 
\gls{kappa} for the sphere-like scenario increases, therefore reducing the 
region in which it is observed.
Flutter is observed neither for $\gls{chi}=2$ nor $3$, while tumbling starts
to appear for heavy particles with large density ratios.
It should be noted that for $\gls{chi}=3$ the first bifurcation leading to the
sphere-like scenario is subcritical, allowing the coexistence of steady vertical
and tumbling modes.
For $\gls{chi}=4$ and $5$ there is no sphere-like scenario.
Flutter is recovered for particles with intermediate non-dimensionalized mass.
Regarding tumbling modes, the subcritical behavior at large density ratios
is progressively increased.
Finally, if the aspect ratio is increased to $\gls{chi}=6$ or $10$ the 
sphere-like scenario is recovered for $\gls{kappa}\le 3$ or $5$, respectively, 
which correspond in both cases to $\gls{mstar} \lesssim 0.25$.
For these aspect ratios the regions in which flutter and tumbling are observed
continue to increase, showing stronger subcriticality in their first 
bifurcation.
It should be noted that although the increase in the region where flutter is 
observed includes light particles, tumbling is restricted to heavy particles.
It should also be mentioned that for large aspect ratios $\gls{chi}\ge 5$, 
intermittent modes can be found between the flutter and tumbling regions, 
especially (but not exclusively) in the non-overlapping space between these two.

\subsection{Geometric definitions and notation}             
\label{sec:definitions}

In this section we introduce the notation used to present the subsequent 
results.
When the regime under study is axisymmetric with respect to the axis 
$z^\text{fix}$ (regime A), no additional definitions are needed. 
However, when the axisymmetry is broken, we define the following local 
coordinate system as in \cite{uhlmann:2014b}.

First, we define a reference system that maintains one coordinate along the 
vertical direction \gls{ez}, while the horizontal directions are given by the 
projection of the particle velocity and the direction perpendicular to it, viz.: 
\begin{subequations}
\begin{align}
\gls{epH}&=\left(\gls{vup_x},\gls{vup_y},0\right)/%
          \sqrt{{\gls{vup_x}}^2+{\gls{vup_y}}^2} , \label{eq:epH} \\
\gls{epHzp}&=\gls{ez}\times\gls{epH}.              \label{eq:epHzp}
\end{align}
\end{subequations}
Note that the temporal dependence in cases with unsteady particle motion will be
made precise below.
With these definitions the following linear and angular velocity projections, 
made dimensionless with the gravitational velocity \gls{Ug}, can be defined:
\begin{subequations}
\begin{align}
\gls{vup_h}    &=\left(\gls{vup}/\gls{Ug}\right)\cdot\gls{epH}  ,\label{eq:vup_h}\\
\gls{vup_hzp}  &=\left(\gls{vup}/\gls{Ug}\right)\cdot\gls{epHzp},\label{eq:vup_hzp}\\
\gls{vup_v}    &=\left(\gls{vup}/\gls{Ug}\right)\cdot\gls{ez}   ,\label{eq:vup_v}\\
\gls{vomep_h}  &=\left(\gls{vomep}\gls{dd}/\gls{Ug}\right)\cdot\gls{epH}  ,\label{eq:vup_h}\\
\gls{vomep_hzp}&=\left(\gls{vomep}\gls{dd}/\gls{Ug}\right)\cdot\gls{epHzp},\label{eq:vup_hzp}\\
\gls{vomep_v}  &=\left(\gls{vomep}\gls{dd}/\gls{Ug}\right)\cdot\gls{ez} .  \label{eq:vup_v}
\end{align}
\end{subequations}
The particle Reynolds is defined as follows
\begin{equation}\label{eq:Re_ll}
\gls{Re_ll}=\frac{\left|\gls{vup}\right|\gls{dd}}{\gls{nu}}
 =\frac{\left|\gls{vup}\right|}{\gls{Ug}}\gls{Ga}.
\end{equation}
Please note that \gls{vup} is defined with respect to the fixed coordinate 
system \gls{Ofix} (see figure \ref{fig:problem_description}), such that it 
automatically constitutes a relative velocity with respect to the ambient 
fluid \citep[this is different from the definition used in ][]{uhlmann:2014b}.

\begin{figure} 
\makebox[\textwidth][c]{ 
\includegraphics[scale=1.0]{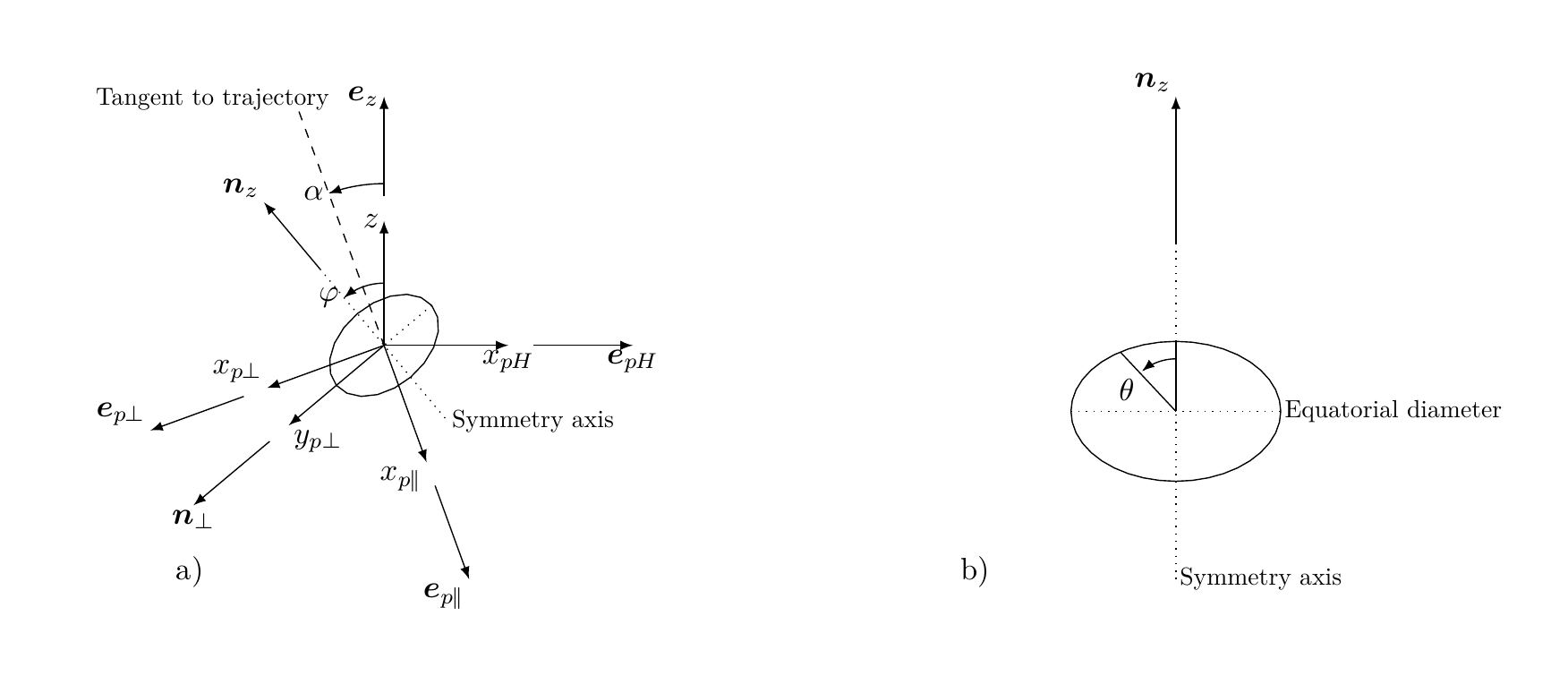} 
}
\caption{
Sketch of a) the coordinate system defined from the trajectory of the 
particle in the plane defined by the unitary vectors \gls{epH} and \gls{ez}
and b) the parametric variable $\theta$ used to represent 
the pressure distribution.
\label{fig:geometric_defs}} 
\end{figure}

Second, we define a reference system that shares the horizontal component 
\gls{epHzp}, which is normal to the trajectory plane.
The unit vector aligned with the particle's trajectory is
\begin{equation}\label{eq:epll}
\gls{epll} = \gls{vup}/\left|\gls{vup}\right|,
\end{equation}
and the unit vector perpendicular to \gls{epHzp} and \gls{epll} (see 
figure~\ref{fig:geometric_defs}a) is obtained from:
\begin{equation}\label{eq:epp}
\gls{epp}=\gls{epHzp}\times\gls{epll}.
\end{equation}
With this, we can define the following components of the relative fluid velocity
$\gls{vur}(\gls{vx},t)=\left(\gls{vur_x},\gls{vur_y},\gls{vur_z}\right)=%
\gls{vu}-\gls{vup}$
\begin{subequations}\label{eq:vur}
\begin{align}
\gls{vur_ll}   &=\left(\gls{vur}/\gls{Ug}\right)\cdot(-\gls{epll})  , \\
\gls{vur_p}    &=\left(\gls{vur}/\gls{Ug}\right)\cdot\gls{epp}  , \\
\gls{vur_Hzp}  &=\left(\gls{vur}/\gls{Ug}\right)\cdot\gls{epHzp} .
\end{align}
\end{subequations}
Note that in the case of axi-symmetric solutions the axial component of the 
relative velocity is simply \gls{vur_ll} and the radial component is denoted as
$\gls{vur_r}=\sqrt{\gls{vur_x}^2+\gls{vur_y}^2}/\gls{Ug}$.

Finally, we define the direction $y_{p\perp}$ along the unitary vector
$\vec{n}_{\perp}$, which is perpendicular to the symmetry axis and contained in the trajectory plane
\begin{equation}
\vec{n}_{\perp}=\vec{n}_{z} \times \gls{epHzp} \,,
\end{equation}
where $\vec{n}_{z}$ is a unitary vector parallel to the symmetry axis of the spheroid
pointing towards positive values of $z$.

\section{Spectral/spectral--element computations}
\label{sec:sem}

\subsection{Numerical method}
\label{sec:sem/method}

\begin{figure} 
\makebox[\textwidth][c]{ 
\includegraphics[scale=1.0]{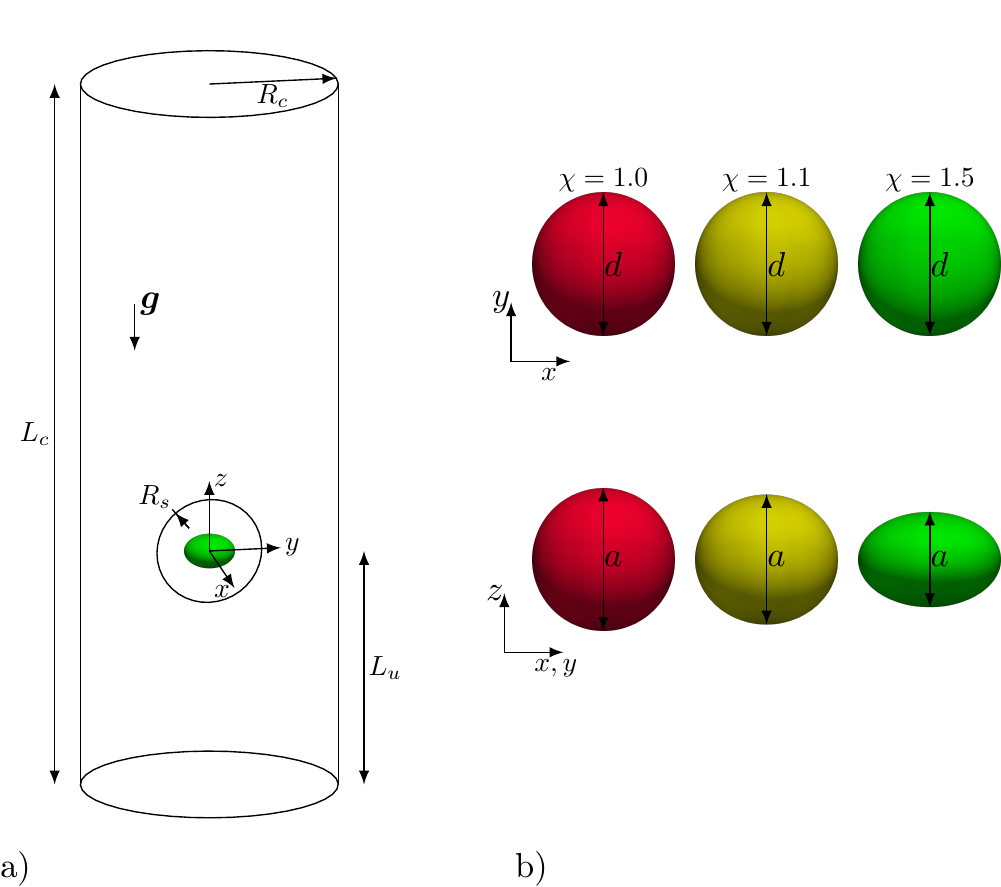} 
}
\caption{a) Cylindrical domain used in the \gls{sem} computations.
The length of the domain is $L_c=15\gls{dd}$ and its radius $R_c=2.67\gls{dd}$.
The  spherical sub-domain of radius $R_s=\gls{dd}$ is also represented.
c) Top and side view of the particles used in the simulation ($\gls{chi}=
1.1,1.5$) together with a sphere ($\gls{chi}=1$) to appreciate the 
differences.
\label{fig:sem_setup}} 
\end{figure}

The reference data has been generated with the aid of the
spectral/spectral-element method proposed by \cite{chrust:2013}. 
It solves the Navier-Stokes equations
(\ref{eq:gov})
in a cylindrical domain of length $L_c=15d$ and radius $R_c=2.67d$, 
with the cylinder axis along the gravity vector (cf.\
figure~\ref{fig:sem_setup}).
The particle's centroid is located on the cylinder axis at a distance
$L_u=5d$ from the lower (upstream) domain boundary. 
A spherical sub-domain with a radius equal to $R_s=d$ is fitted
around the spheroidal particle. The sub-domain is rotating with the
angular particle motion, while the outer computational domain is
linearly translating with the particle motion. 
As a consequence, the mesh remains body-conforming, while the
outer computational domain maintains its vertical orientation. 

%
At the inflow (bottom) cylinder basis the velocity is set equal to
zero to simulate an asymptotically quiescent fluid. At the outflow
(top) cylinder basis and at its side a no stress Neumann boundary
condition is imposed on the velocity field and a zero pressure is set.

\begin{figure}
\begin{center}
\includegraphics{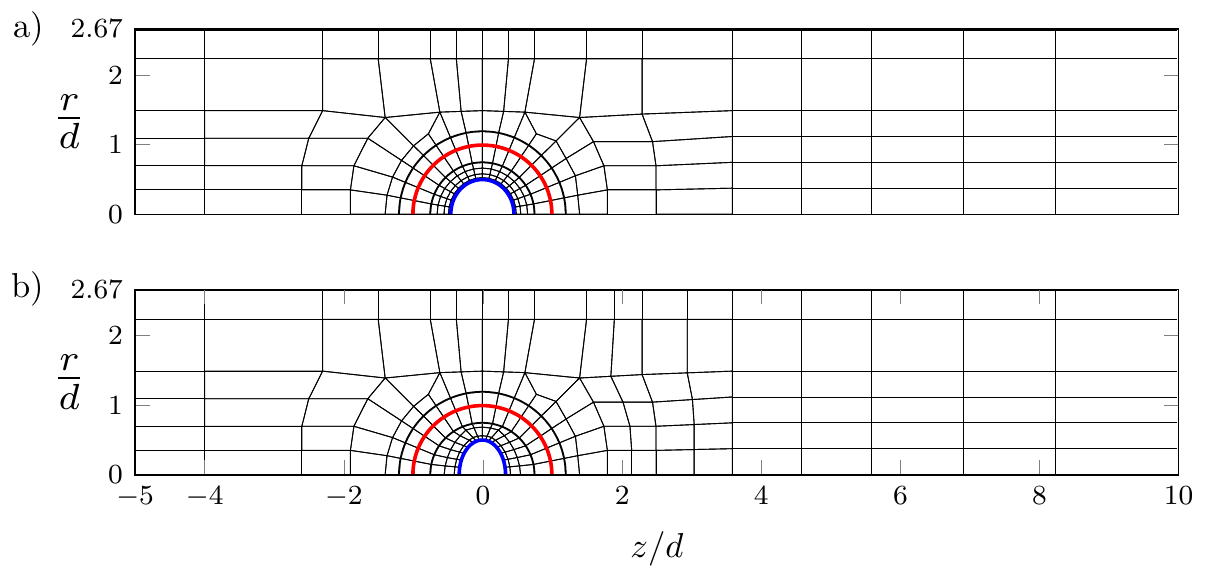}
\caption{Spectral element grid used for the spheroid with a) $\gls{chi}=1.1$
and b) $\gls{chi}=1.5$.
The radial coordinate of the reference cylindrical system is $r$ and the axial
$z$.
The surface of the spheroids is represented in blue and the interface between inner and outer
subdomains in red.
\label{fig:semgrid}}
\end{center}
\end{figure}

The Navier-Stokes equations are spatially discretized in the axial/radial plane
with the aid of the spectral-element method, while a truncated Fourier expansion
is employed in the azimuthal direction. 
In the present work the Fourier expansion is truncated at the $15$-th mode, and
the spectral-element meshes shown in figure~\ref{fig:semgrid} are used with a 
polynomial order of $6$.
The solutions of the inner sub-domain and the outer domain are
coupled at the common spherical interface through spherical harmonics
\citep[please refer to][for more details and validation]{chrust:2012}. 
Grid convergence for this choice as well as the location of the interface
between the two subdomains
\review{has been demonstrated previously in the work of Chrust et
  al. (2012)}{%
  has been demonstrated previously both for spheres and thin disks
  in \S~3.12 of the PhD thesis by
  \cite{chrust:2012} and in Tables~I and II of \cite{chrust:2013}.
}
%
The temporal discretization is performed with a third-order Adams-Bashforth 
scheme. 
The present numerical method has been used in the past for the simulation of
the sedimentation of spheres \citep{jenny:2004,uhlmann:2014b}, disks 
\citep{chrust:2013} as well as spheroids \citep{zhou:2017}.
Note that, in order to reduce the computational cost of these simulations for 
further use for validation purposes, the domain used in this work is smaller 
than the one presented in \cite{chrust:2012,chrust:2013,zhou:2017}.
In \cite{uhlmann:2014b} the difference in the results using both of these domain
sizes was shown to be small for the case of spheres at comparable Galileo 
numbers.
%

\subsection{Results}
\label{sec:sem/results}

\begin{figure}
\begin{center}
\makebox[\textwidth][c]{
\includegraphics{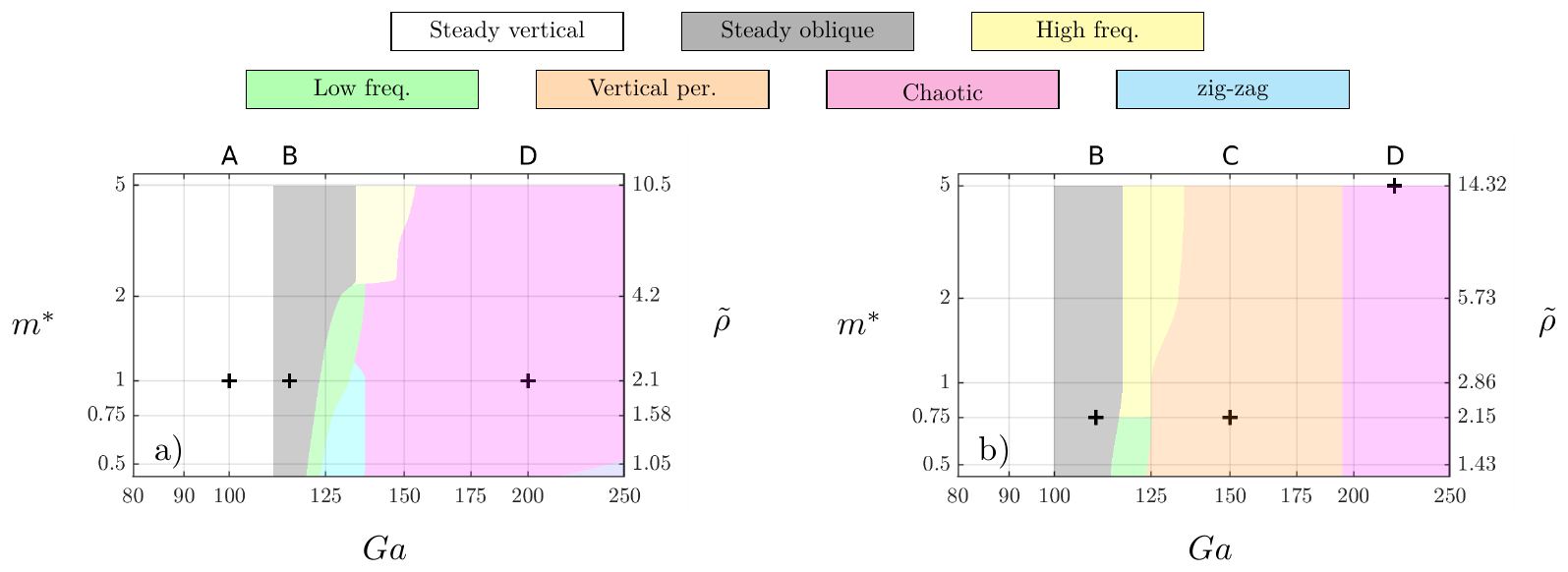}}
\caption{Flow regime maps for a) $\gls{chi}=1.1$ \citep{zhou:2017} and 
b) $\gls{chi}=1.5$ (unpublished).
\label{fig:regimesdetail}}
\end{center}
\end{figure}

In this section we present detailed results of selected cases for low 
($\gls{chi}=1.1$) and moderate ($\gls{chi}=1.5$) aspect ratio oblate spheroids.

We have selected four different regimes to analyze in detail and provide 
benchmark data to use for validation purposes.
These regimes are the steady axisymmetric (labelled A), the steady 
oblique (labelled B), the periodic oscillating vertical in the mean (labelled C)
and the chaotic regime (labelled D).
The complete set of cases can be found in table \ref{tab:cases} and the 
representation of these cases in the $\gls{Ga},\gls{kappa}$ diagram in figure
\ref{fig:regimesdetail}.
Both the zig-zag and vertical periodic regimes refer to periodically 
oscillating trajectories, vertical in the mean. 
The zig-zag regime is part of the sphere-like scenario and has about four
times longer period than the vertical periodic regimes in figure
\ref{fig:regimesdetail}b.

\begin{table}
\begin{center}
\caption{Set of cases indicating the flow regime and the parameter point
${(\gls{chi}, \gls{Ga}, \gls{kappa})}$. The non-dimensional mass ${\gls{mstar}=%
\gls{kappa}\pi/\left(6\gls{chi}\right)}$ is also included for completeness.
\label{tab:cases}}
\begin {tabular}{cccccc}%
\toprule Case identifier&Flow regime&$\gls {chi}$&$\gls {Ga}$&$\gls {mstar}$&$\gls {kappa}$\\\toprule %
\gls {A11M100}&Steady vertical&\ensuremath {1.1}&\ensuremath {100}&\ensuremath {1}&\ensuremath {2.1}\\\hline %
\gls {B11M100}&Steady oblique&\ensuremath {1.1}&\ensuremath {115}&\ensuremath {1}&\ensuremath {2.1}\\%
\gls {B15M075}&&\ensuremath {1.5}&\ensuremath {110}&\ensuremath {0.75}&\ensuremath {2.14}\\\hline %
\gls {C15M075}&Vertical periodic&\ensuremath {1.5}&\ensuremath {150}&\ensuremath {0.75}&\ensuremath {2.14}\\\hline %
\gls {D11M100}&Chaotic&\ensuremath {1.1}&\ensuremath {200}&\ensuremath {1}&\ensuremath {2.1}\\%
\gls {D15M500}&&\ensuremath {1.5}&\ensuremath {220}&\ensuremath {5}&\ensuremath {14.32}\\\toprule %
\end {tabular}%
\end{center}
\end{table}

\subsubsection{Steady vertical regime}
\label{sec:sem/results/A}

For Galileo number $Ga=100$ an oblate spheroid of aspect ratio $\gls{chi}=1.1$ 
reaches a steady axisymmetric state.
The symmetry axis of the solution coincides with the symmetry axis of the 
spheroid, which in turn is aligned with the gravitational acceleration vector.
No particle rotation is observed; therefore, the particle kinematics are
uniquely defined by the vertical component of the velocity $\gls{vup_v}$ (see 
table \ref{tab:results_refA}).

\begin{figure}
\makebox[\textwidth][c]{
\includegraphics[]{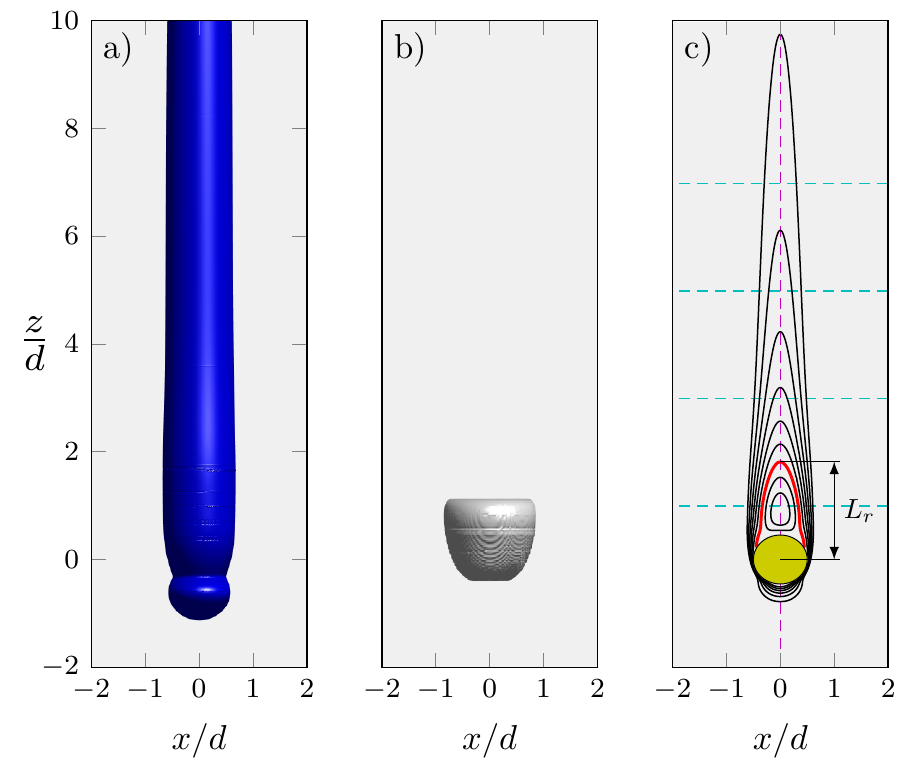}
}
\caption{Flow visualization for a spheroid with $\gls{chi}=1.1$ at $\gls{Ga}=100$,
$\gls{kappa}=2.1$ ($\gls{mstar}=1$) which results in a steady vertical 
regime.
Isosurface of a) relative velocity $\gls{vur_ll}=1.5$ and b) $\gls{l2}=-0.015%
\gls{l2ref}$ \citep{jeong:1995}.
c) Isocontours of relative velocity $\gls{vur_ll}=(-0.4:0.2:1.4)$ in a vertical plane
passing through the center of the particle.
The red line in c) corresponds to $\gls{vur_ll}=0$.
The cyan and purple dashed lines indicate the location of the velocity profiles
used for benchmarking purposes in \S~\ref{sec:ibm/results/A} and given in the supplementary material.
The cross stream profiles (cyan lines) are located at $z/d=1,3,5,7$.
\label{fig:visu_SV}}
\end{figure}

Figure \ref{fig:visu_SV} shows different flow visualizations of spectral-element
data interpolated to a Cartesian grid for later comparison.
As expected by the axi-symmetry found in this case, the wake is aligned with
the trajectory of the particle (panel a).
Following \cite{jeong:1995}, vortical structures are identified by regions 
\review{where $\gls{l2}<0$, which is the second largest eigenvalue of the tensor  
$\gls{graduS}^2+\gls{graduO}^2$}{%
where the second largest eigenvalue of the tensor $\gls{l2}=\gls{graduS}^2+\gls{graduO}^2$ is negative
} (where \gls{graduS} and \gls{graduO} are the 
symmetrical and anti-symmetrical parts of the fluid velocity gradient tensor), 
showing a clear toroidal vortex around the particle (panel b).
The flow is also characterized by a recirculation region attached to the 
downstream face of the particle and an abrupt deceleration close to the 
stagnation point in the upstream face of the particle (panel c).
Figure \ref{fig:visu_SV}c also shows the recirculation length \gls{Lr}, which is
defined as the largest distance between any point on the boundary of the 
recirculation region ($\gls{vur_ll}=0$) and the center of the particle.
The value of \gls{Lr} for the steady vertical case is presented in table 
\ref{tab:results_refA}.

The pressure field is obviously also axisymmetric in this case. 
Pressure profiles along the great circles will be discussed below in section
\ref{sec:ibm/results/A}, cf.~figure~\ref{fig:greatcircle_SV}.

\begin{table}
\begin{center}
\caption{
\gls{sem} reference results for steady vertical regime of spheroid with 
$\gls{chi}=1.1$.
\label{tab:results_refA}}
\begin {tabular}{cccc}%
\toprule Case&\gls {vup_v}&$\gls {Lr}/\gls {dd}$&$\gls {Re_ll}$\\\toprule %
\gls {A11M100}&\ensuremath {-1.6863}&\ensuremath {1.818}&\ensuremath {168.63}\\\toprule %
\end {tabular}%
\end{center}
\end{table}

\subsubsection{Steady oblique regime}
\label{sec:sem/results/B}

The second regime under study is still steady, but instead of axisymmetric, the
solution is planar symmetric.
Both aspect ratios are evaluated ($\gls{chi}=1.1, 1.5$) with Galileo $100$ and 
$110$, respectively.
In this regime the kinematics of the particle are defined by its vertical 
(\gls{vup_v}) and horizontal (\gls{vup_h}) velocity, and by the angle 
\gls{alpha_p} between the symmetry axis of the spheroid and the vertical (cf. 
the sketch in figure~\ref{fig:geometric_defs}).
We also include the angle of the particle's gravity center trajectory with 
respect to the  vertical
\begin{equation}
\tan\left(\alpha\right)=\frac{\gls{vup_h}}{\left|\gls{vup_v}\right|}.
\end{equation}
Differently than for spheres, the angular velocity of the converged state is 
zero \citep{jenny:2004,uhlmann:2014b}.
Note that the orientation of the vector \gls{epH} is arbitrarily selected by the
solution.
The resulting values obtained for both aspect ratios can be seen in table 
\ref{tab:results_refB}.

\begin{table}
\begin{center}
\caption{\gls{sem} reference results for steady oblique regime of spheroid with 
$\gls{chi}=1.1$ and $\gls{chi}=1.5$. 
\label{tab:results_refB}}
\begin {tabular}{cccccccc}%
\toprule Case&$\gls {chi}$&\gls {vup_v}&\gls {vup_h}&$\gls {alpha_t}(^\circ )$&$\gls {alpha_p}(^\circ )$&$\gls {Lr}/\gls {dd}$&$\gls {Re_ll}$\\\toprule %
\gls {B11M100}&\ensuremath {1.1}&\ensuremath {-1.76}&\ensuremath {0.057}&\ensuremath {1.858}&\ensuremath {2.755}&\ensuremath {1.966}&\ensuremath {202.46}\\%
\gls {B15M075}&\ensuremath {1.5}&\ensuremath {-1.682}&\ensuremath {0.113}&\ensuremath {3.842}&\ensuremath {5.318}&\ensuremath {2.039}&\ensuremath {185.43}\\\toprule %
\end {tabular}%
\end{center}
\end{table}

\begin{figure}
\begin{center}
\makebox[\textwidth][c]{
\includegraphics[scale=0.9]{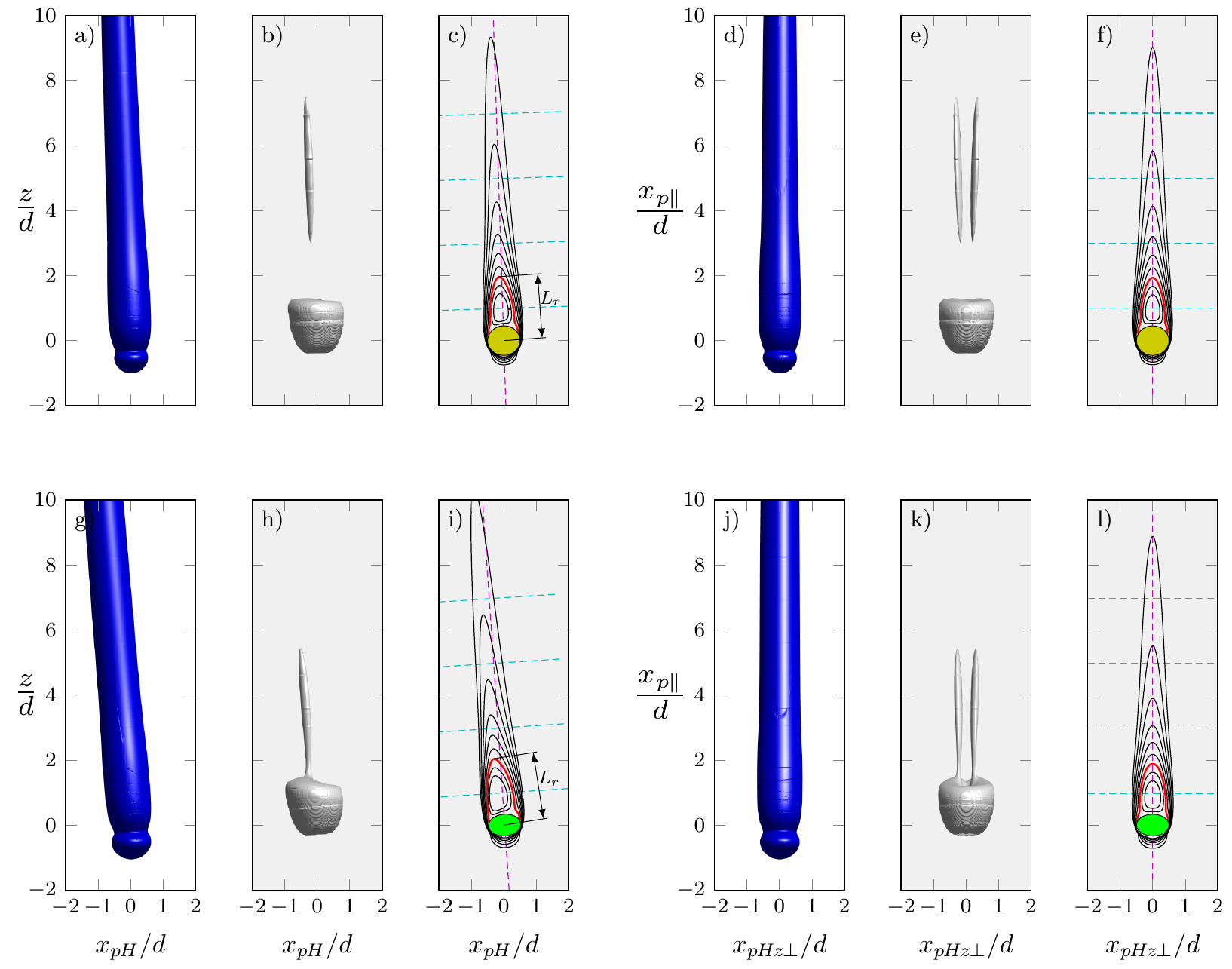}
}
\caption{Flow visualization of cases \gls{B11M075} (top row) and \gls{B15M075}
(bottom row).
Panels a, d, g and j show isosurface of $\gls{vur_ll}=1.5\gls{Ug}$, panels b and
e, show isosurfaces of $\gls{l2}=-10^{-4}\gls{l2ref}$ and panels h and k, 
$\gls{l2}=-0.015\gls{l2ref}$.
Panels c, f, i and l show isocontours of $\gls{vur_ll}=-0.4:0.2:1.2$, 
highlighting with a thick, red line the isocontour $\gls{vur_ll}=0$.
The cyan and purple dashed lines in panels c,f,i,l indicate the location of 
the velocity profiles provided for benchmarking purposes in the supplementary material.
The cross stream profiles (cyan lines) are located at $\gls{xpll}/d=-1,-3,-5,-7$.
\label{fig:visu_SO}}
\end{center}
\end{figure}

Figure \ref{fig:visu_SO} shows side and front views of the isocontours of 
relative velocity ($\gls{vur_ll}=1.5$) and of $\gls{l2}$ for the spheroid 
with $\gls{chi}=1.1 (1.5)$ and $\gls{Ga}=100\,(110)$.
The wake shows a similar structure to that found in the steady regime A, but it
is here aligned with the particle's oblique trajectory.
It can also be seen that the angle of the trajectory followed by the particle
with $\gls{chi}=1.1$ (\ref{fig:visu_SO}a) is smaller than that of the particle
with $\gls{chi}=1.5$ (\ref{fig:visu_SO}g).
Regarding the vortical structure, the strong toroidal vortex is still present 
and, additionally, two counter rotating streamwise vortices appear a few 
diameters downstream of the particle.
Note that these double-threaded vortices are very weak compared to the toroidal
vortex, especially for the spheroid with $\gls{chi}=1.1$.
Therefore, different thresholds of \gls{l2} are used in these visualizations.

Figure \ref{fig:visu_SO} also shows isolines of constant streamwise relative
velocity \gls{vur_ll}.
The side view of the contours of \gls{vur_ll} show that the furthest point of 
the recirculation bubble, which is used to measure the recirculation length
\gls{Lr}, is not located on the trajectory of the center of the particle,
but slightly deviated in the direction \gls{epp}.
The planar symmetry of this regime can be seen in the frontal view.
Also indicated in the figure are the downstream locations where cross-stream 
profiles of the velocity components are reported in the supplementary material; 
these are also used in the benchmarking of the \gls{ibm} results in 
section~\ref{sec:ibm/results/B}

\begin{figure}
\begin{center}
\includegraphics[scale=1.0]{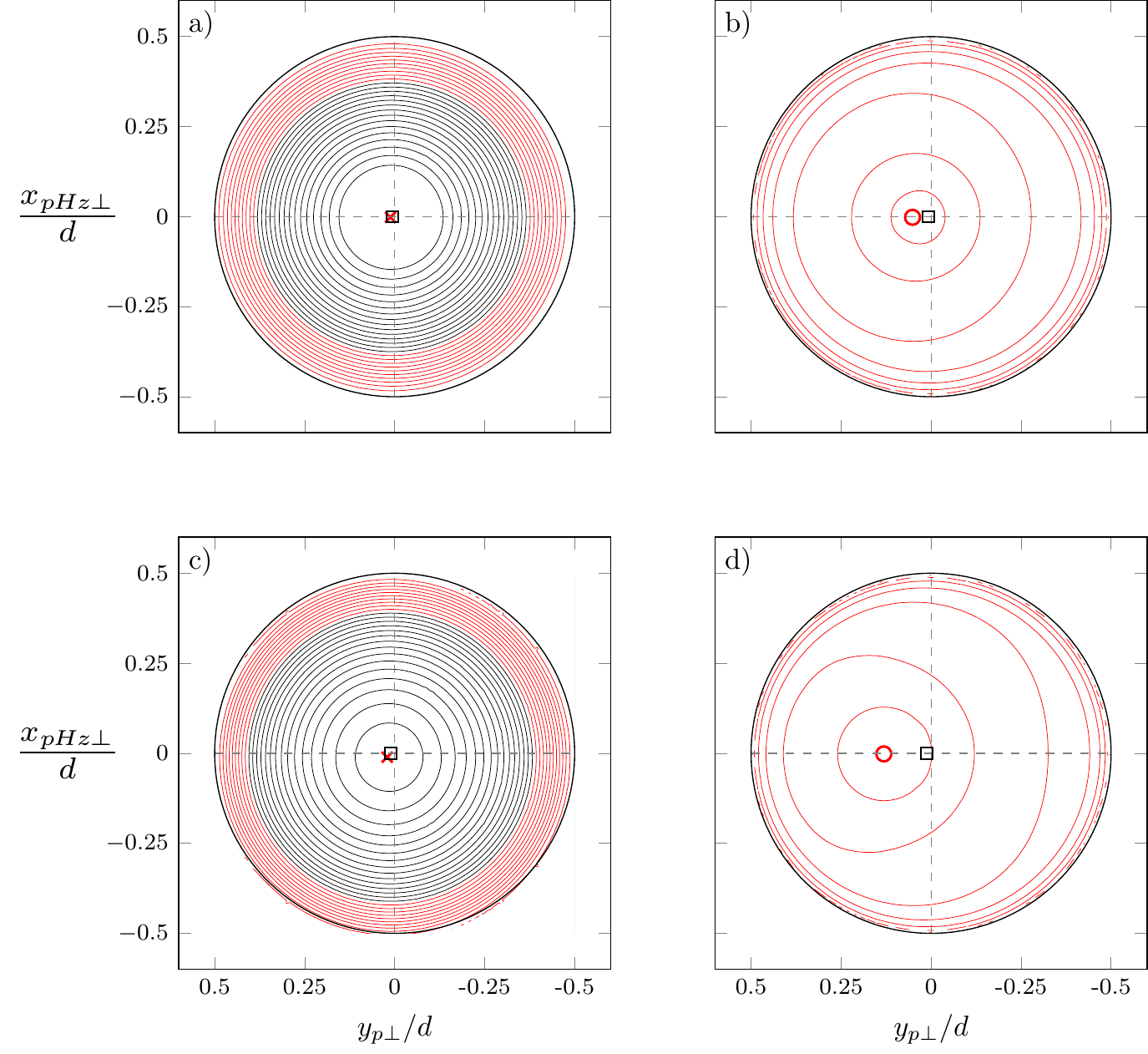}
\caption{%
Isocontours of pressure on \review{ a) upper and b) lower surface of case 
\gls{B11M075} and c) upper and d) lower surface of \gls{B15M075} perpendicular
to the symmetry axis.}{the surface of a,b) \gls{B11M075} and c,d) \gls{B15M075}.}
\review{}{
Left (right) panels correspond to the projection of the contours of the lower
(upper) surface on a plane perpendicular to the symmetry axis.
}
Positive contours are represented in black ($0.1:0.1:1.5$) and
zero and negative contours are represented in red ($-1.5:0.1:0$).
The red circle (cross) marker in panels a and c (b and d) indicates 
the position of the minimum (maximum) pressure.
The intersecting point of the particle's center trajectory with the 
surface of the spheroid is represented with a square in every panel. 
\label{fig:press_SO}}
\end{center}
\end{figure}

Figure \ref{fig:press_SO} shows isocontours of constant pressure on the %
\review{%
top (or downstream) and bottom (or upstream) surfaces in}{surface of} both 
spheroids with $\gls{chi}=1.1$ and $1.5$ for the steady oblique regime.
\review{%
The contours on the surface are projected to a plane perpendicular to the symmetry
axis of the spheroid which, in these cases, coincides with a plane spanned by the
vectors \gls{epp} and \gls{epHzp}
(please recall that the plane spanned by 
\gls{epp} and \gls{epHzp} depends only on the trajectory of the particle and 
will not, in general, coincide with the plane perpendicular to the symmetry axis
of the spheroid).
}{%
The contours on the surface are projected to a plane perpendicular to the symmetry
axis of the spheroid which is not, in general, parallel to the trajectory of the
particle.
Therefore, the projection plane is spanned by $\vec{n}_\perp$ and \gls{epHzp} 
(see figure \ref{fig:geometric_defs}a).
We identify as upper (lower) the surface that predominatly faces downstream or
upwards (upstream or downwards).
}
Interestingly the distribution of the contours in the upstream-facing side of
the spheroid (panels a and b) is almost concentrical, as it would be in the case
of an axi-symmetric solution.
Only the small displacement of the pressure maxima in the direction $\gls{epp}$
breaks the axi-symmetry of the contours.
Conversely, the contours in the downstream face of the particle (panels a and c) 
are clearly not axi-symmetric, with a more pronounced deviation in the case with 
higher aspect ratio (panel d).
In the future it will be of interest to investigate in more detail the trends 
(with the aspect ratio \gls{chi}) of the flow pattern in the wake, and of the 
resulting surface stress distribution.

\subsubsection{Vertical periodic regime}
\label{sec:sem/results/C}

In this section we evaluate the periodic oscillations that appear for
a spheroid of aspect ratio $\gls{chi}=1.5$, $\gls{Ga}=150$ and $\gls{kappa}=2.14$ 
($\gls{mstar}=0.75$) (see figure \ref{fig:regimesdetail}b).
The motion in this regime is vertical in the mean and it is restricted
to a plane.
Therefore, the kinematics can be reduced to vertical (\gls{vup_v}) and
horizontal (\gls{vup_h}) velocity components and to the angular velocity around the
axis perpendicular to the plane in which the motion takes place (\gls{vomep_hzp}).
It should be noted that the definition of \gls{epH} \eqref{eq:epH} involves 
quantities evaluated at the reference time \gls{ts},
which is defined as the time instant of the oscillation cycle in which
$\sqrt{\gls{vup_x}^2+\gls{vup_y}^2}$ is maximum.
Therefore, equation \eqref{eq:epH} is rewritten as
\begin{equation}\label{eq:epHs}
\gls{epH}=\left(\gls{vup_x}(\gls{ts}),\gls{vup_y}(\gls{ts}),0\right)/%
          \sqrt{{\gls{vup_x}}^2(\gls{ts})+{\gls{vup_y}}^2(\gls{ts})}.
\end{equation}
The remaining definitions presented in \S~\ref{sec:definitions} prevail.

\begin{figure} 
\makebox[\textwidth][c]{ 
\includegraphics[scale=1.0]{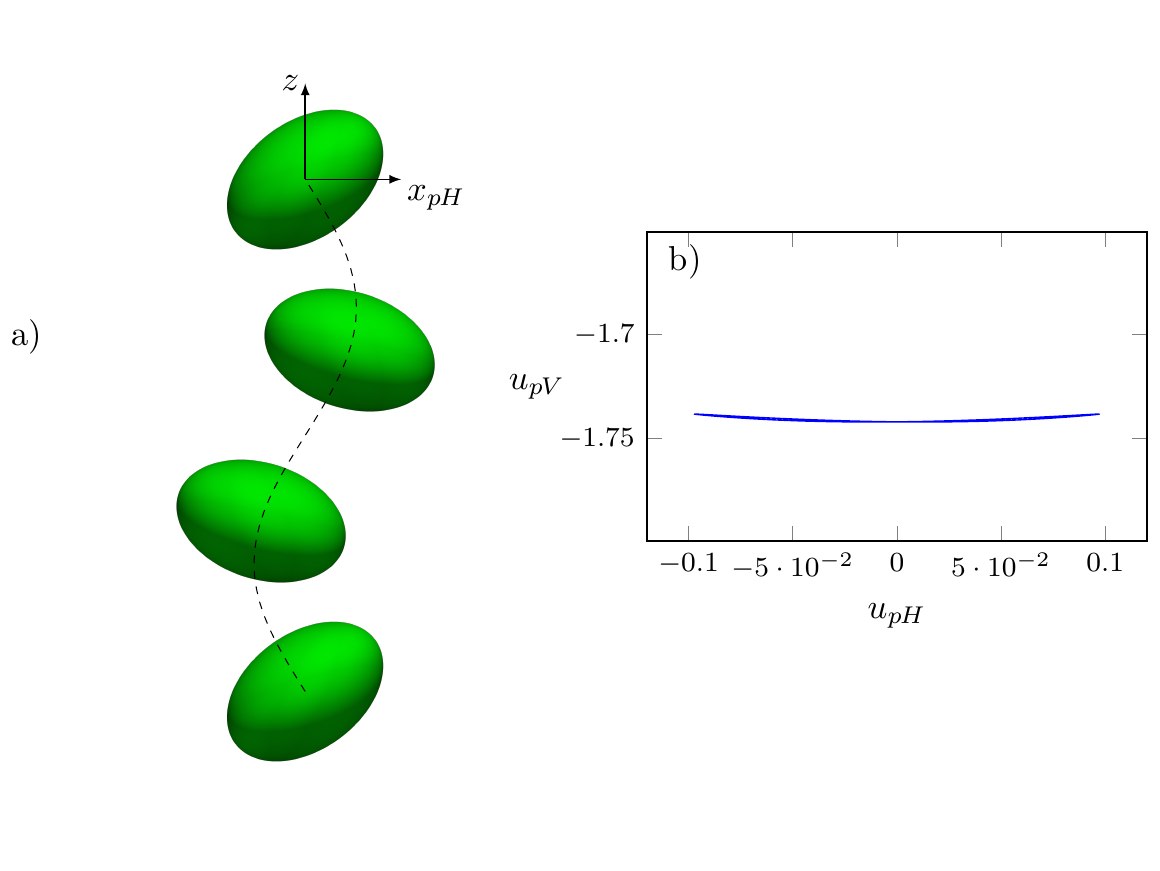} 
}
\caption{a) Sketch of the kinematics in the vertical periodic regime. Note that 
the inclination and path deviations are exaggerated for clarity.
b) Phase-space trajectory in the plane defined by the horizontal and vertical
velocities \gls{vup_h} and \gls{vup_v} for the periodic oscillating case of the 
spheroid with $\gls{chi}=1.5$, $\gls{Ga}=150$ and $\gls{mstar}=0.75$ ($\gls{kappa}=2.14$)
computed with the \gls{sem}.
\label{fig:trajectories_OSCILLATORY}} 
\end{figure}

Table \ref{tab:results_refC} shows the results of the vertical periodic case.
These cases are time periodic with frequency of oscillation \gls{fosc} and
period  \gls{Tosc}, from which we can define the Strouhal number $\gls{St}=
\gls{fosc}\gls{dd}/\gls{Ug}$.
The amplitude and mean values of any function $\phi$ can be obtained from the 
fully developed data as follows:
\begin{subequations}
\begin{align}
\phi'& =\mathrm{max}_t\left( \phi(t) \right) -
      \mathrm{min}_t\left( \phi(t) \right), \quad \gls{ts}<t<\gls{ts}+\gls{Tosc} ,\\
\overline{\phi}& =\frac{1}{\gls{Tosc}}\int_{\gls{ts}}^{{\gls{ts}+\gls{Tosc}}} \phi(t) \wrt t.
\end{align}
\end{subequations}
The time history of the kinematic variables follows a sinusoidal shape (figure 
omitted).
The amplitude of the oscillation in the vertical direction (\gls{vup_v_a}) is 
very small compared to the horizontal one (\gls{vup_h_a}),
which in turn is small compared to the mean vertical velocity of the particle
(\gls{vup_v_m}) (see figure \ref{fig:trajectories_OSCILLATORY}b).
Regarding the orientation of the particle, the axis of the spheroid is 
tilted so that the drag is maximized by aligning the axis of the spheroid with
the trajectory, similarly to the steady oblique regime (see figure
\ref{fig:trajectories_OSCILLATORY}a).
This regime is referred in the literature as a fluid mode \citep{zhou:2017}, 
in which the unsteady motion is governed by the fluid behavior: this corresponds
to a planar symmetric wake with alternate shedding of hairpin-like vortices.

\begin{table} 
\caption{\gls{sem} reference results in the vertical periodic regime.
\label{tab:results_refC}} 
\makebox[\textwidth][c]{ 
\begin {tabular}{cccccccc}%
\toprule Case&\gls {St}&\gls {Re_ll}&\gls {vup_v_m}&\gls {vup_v_a}&\gls {vup_h_a}&\gls {vomep_hzp_a}&$\gls {alpha_p_max}(^\circ )$\\\toprule %
\gls {C15M075}&\ensuremath {0.1984}&\ensuremath {261}&\ensuremath {-1.7401}&\ensuremath {0.0038}&\ensuremath {0.1938}&\ensuremath {0.4078}&\ensuremath {9.3055}\\\toprule %
\end {tabular}%
}
\end{table}

\subsubsection{Chaotic regime}
\label{sec:sem/results/D}

\begin{figure}
\begin{center}
\includegraphics{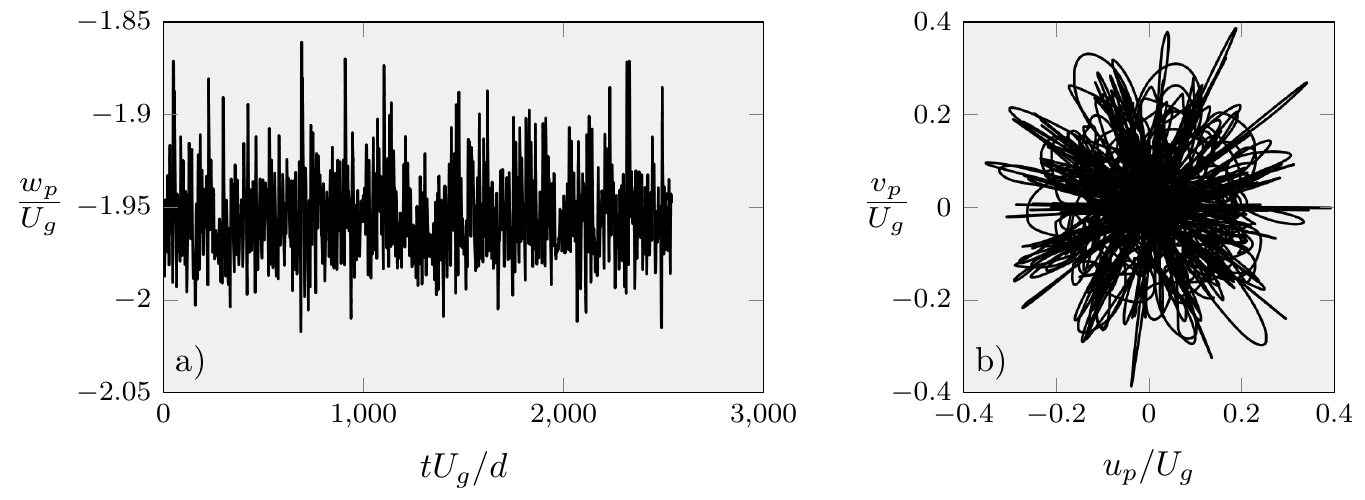}
\caption{Chaotic regime of spheroid with $\gls{chi}=1.1$ at $\gls{Ga}=200$ and
${\gls{mstar}=1}$ (${\gls{kappa}=2.1}$) computed with the \gls{sem}.
a)~Time history of vertical velocity and 
b)~the phase-space trajectory in the plane defined by the two horizontal velocity components. 
\label{fig:sem/chaos}}
\end{center}
\end{figure}

Here, we discuss the data obtained in the regime of chaotic particle motion, 
for $\gls{chi}=1.1$ ($1.5$), $\gls{Ga}=200$ ($220$), and $\gls{kappa}=2.1$ ($14.32$).
Figure \ref{fig:sem/chaos} shows the time history of the vertical velocity
as well as the phase-space plot spanned by the horizontal components of the
velocity for the case with $\gls{chi}=1.5$.
Because of the chaotic behavior of the particle's motion, the results are
described using statistical analysis.
Let the mean of any time dependant variable $\phi(t)$ be defined as 
\begin{equation}\label{eq:mean}
\langle\phi\rangle = \frac{1}{\gls{Tobs}}\int_{\gls{Tobs}} \phi(t)\wrt{}t,
\end{equation}
where \gls{Tobs} is the observation time as stated in table \ref{tab:results_refD}.
Then, the fluctuation of this quantity $\phi''(t)$ is defined as
\begin{equation}\label{eq:fluct}
\phi''(t) = \phi(t) - \langle \phi \rangle.
\end{equation}
%

Table \ref{tab:results_refD} shows results of mean and fluctuating part of the 
linear and angular velocities of both cases.
It can be seen that the amplitude of the vertical velocity fluctuations is very small
compared to the mean value ($<1\%$) and that the fluctuations in the horizontal
direction are approximately $5$ times more intense than in the vertical direction.
There is no mean angular velocity in any of its components.
Regarding the fluctuating part of \gls{vomep},  the intensity of the 
horizontal component is likewise greater than the vertical one, but now by a factor of 
approximately $20$. 
These results are consistent with previous observations in the case of spheres
\citep{uhlmann:2014b}.

\begin{figure}
\begin{center}
\makebox[\textwidth][c]{\includegraphics[scale=0.85]{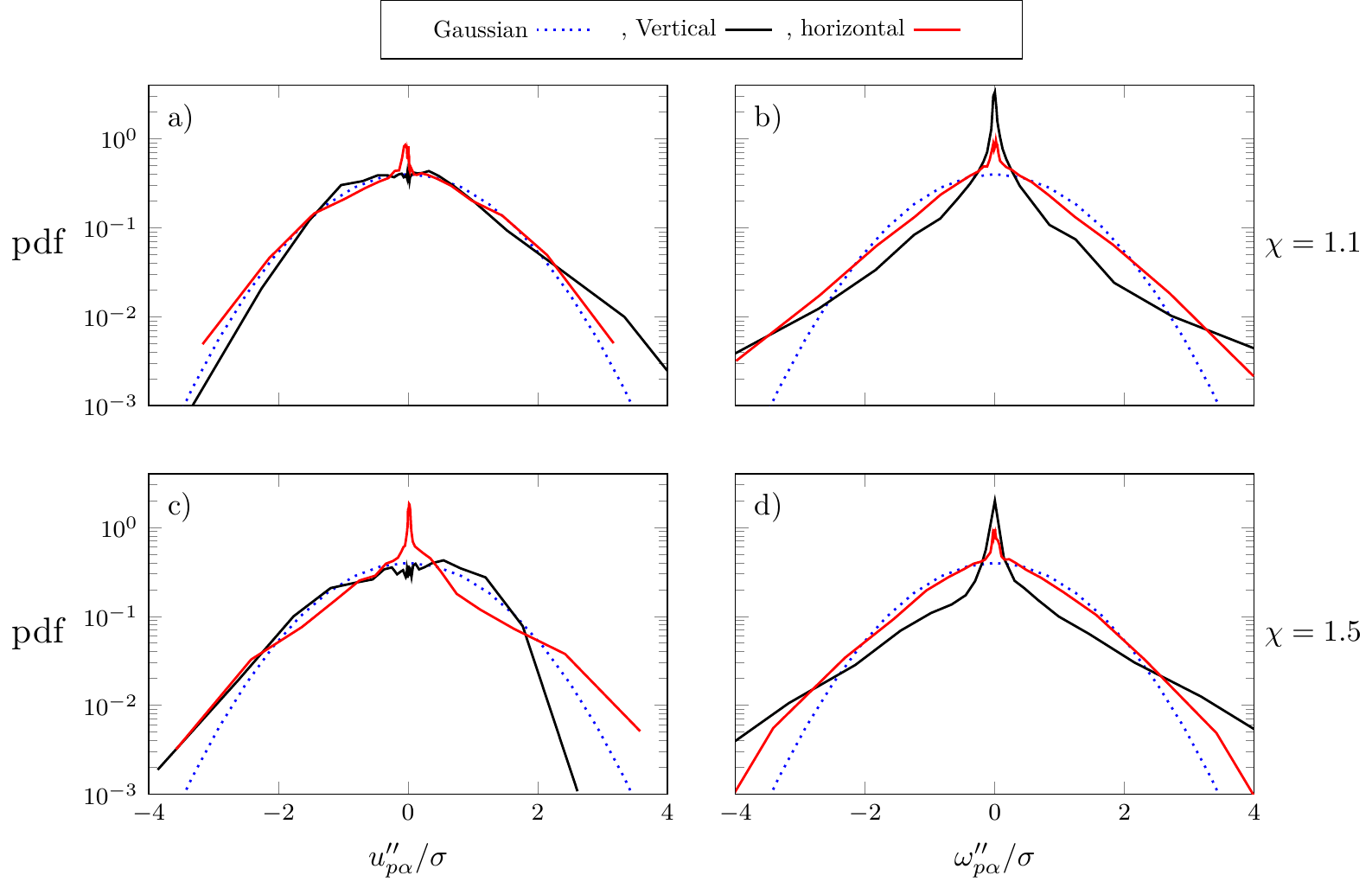}}
\caption{\gls{sem} results in the chaotic regime: \gls{pdf} of linear (left
panels) and angular (right panels) velocities of the spheroid with 
$\gls{chi}=1.1$ and $\gls{kappa}=2.1$ at $\gls{Ga}=200$ (top row) and
$\gls{chi}=1.5$ and $\gls{kappa}=14.32$ at $\gls{Ga}=220$ (bottom row).
The normal distribution is included as a blue line to support the 
graphical interpretation of the data.
\label{fig:chaotic_normpdf_nonuniformbins_ref}}
\end{center}
\end{figure}

Additional information on the kinematics of the particle can be found in the 
representation of the \gls{pdf} for both linear and angular velocities in 
figure \ref{fig:chaotic_normpdf_nonuniformbins_ref}.
For the lower aspect ratio ($\gls{chi}=1.1$) the \gls{pdf} of the horizontal component
is very similar to a normal distribution, but the vertical exhibits some positive
skewness.
The angular velocities of this case show no skewness but moderate excess kurtosis
as compared to a Gaussian.
Interestingly, for $\gls{chi}=1.5$ the \gls{pdf} of the horizontal component of 
the velocity shows positive kurtosis, whereas the vertical component shows negative 
skewness.
Regarding the angular velocities, for $\gls{chi}=1.5$ the horizontal component \gls{pdf} resembles 
that of a normal distribution whereas the vertical component exhibits excess kurtosis.
These results are different from those found for spheres in which the angular velocities
and the vertical velocity approximately follow a normal distribution, and where the horizontal component of
the velocity showed negative excess kurtosis \citep{uhlmann:2014b}.

\begin{table}
\begin{center}
\caption{Averaged quantities of linear and angular velocities for chaotic cases
of spheroids with $\gls{chi}=1.1$ and $\gls{chi}=1.5$ at $\gls{Ga}=200$ and 
$\gls{Ga}=220$ and $\gls{kappa}=2.1$ and $14.32$, respectively. These data were computed with the \gls{sem}.
\label{tab:results_refD}}
\begin {tabular}{ccccccccc}%
\toprule cases&\gls {chi}&$\frac {\gls {vup_z_m}}{\gls {Ug}}$&$\frac {\gls {vup_z_f}}{\gls {Ug}}$&$\frac {\gls {vup_x_f}}{\gls {Ug}}$&$\frac {\gls {vop_z_f}}{\gls {Ug}/\gls {dd}}$&$\frac {\gls {vop_x_f}}{\gls {Ug}/\gls {dd}}$&\gls {Re_ll}&$\gls {Tobs}\gls {Ug}/\gls {dd}$\\\toprule %
\gls {D11M100}&\ensuremath {1.1}&\ensuremath {-1.9533}&\ensuremath {0.0220}&\ensuremath {0.1153}&\ensuremath {0.0025}&\ensuremath {0.0403}&\ensuremath {391}&\ensuremath {2{,}536}\\%
\gls {D15M500}&\ensuremath {1.5}&\ensuremath {-1.8681}&\ensuremath {0.0124}&\ensuremath {0.0464}&\ensuremath {0.0005}&\ensuremath {0.0144}&\ensuremath {411}&\ensuremath {2{,}830}\\\toprule %
\end {tabular}%
\end{center}
\end{table}


\begin{figure} 
\makebox[\textwidth][c]{ 
\includegraphics[scale=1.0]{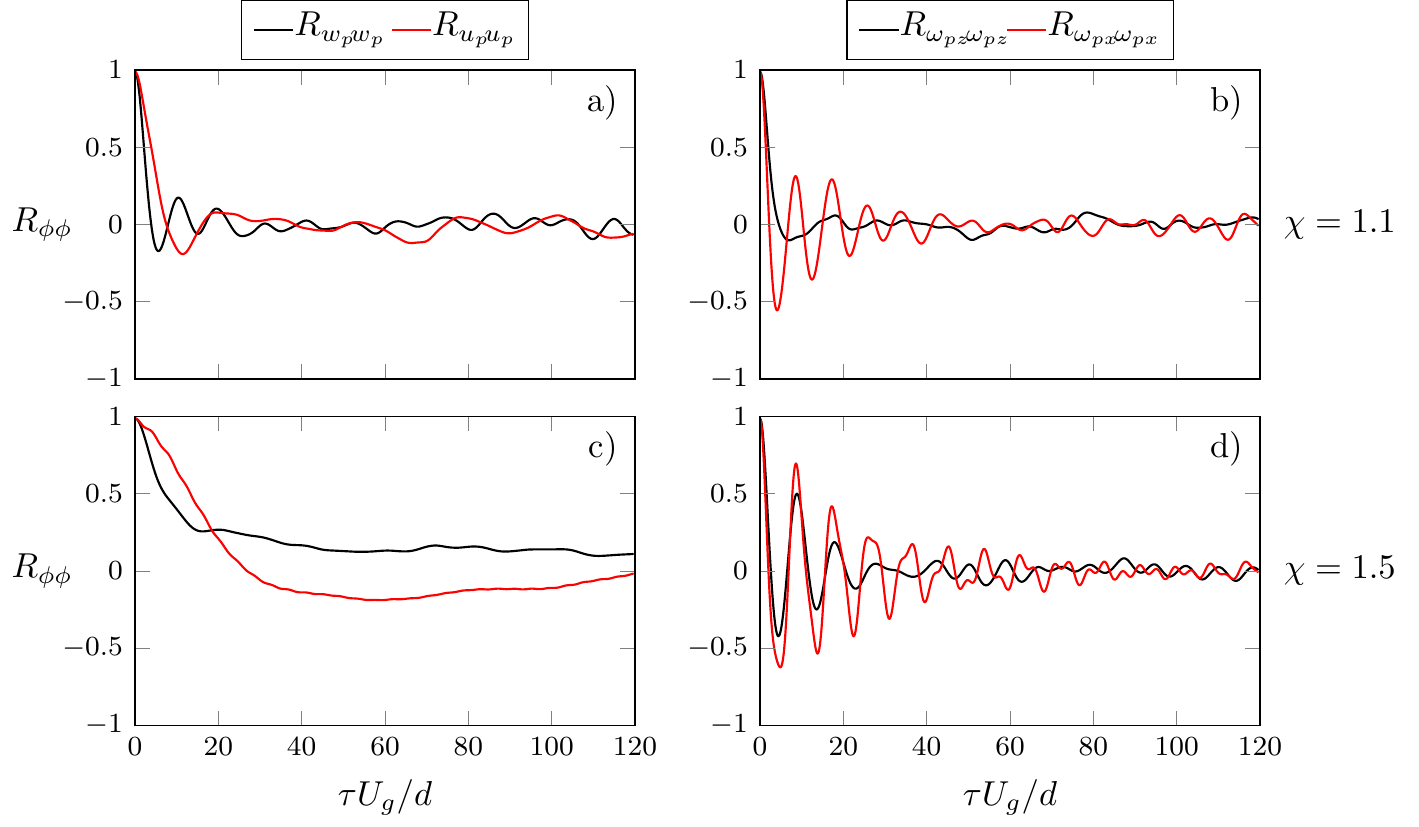} 
}
\caption{\gls{sem} results in the chaotic regime: auto-correlation function of
linear (left) and angular (right) particle velocity for the chaotic regime with
aspect ratio $\gls{chi}=1.1$, $\gls{Ga}=200$, $\gls{kappa}=2.1$ (top row) and 
$\gls{chi}=1.5$, $\gls{Ga}=220$, $\gls{kappa}=14.32$ (bottom row). 
\label{fig:autocorr_D}} 
\end{figure}

The behavior in time of the signals can be evaluated through the auto-correlation
functions of velocity and angular velocity components.
The auto-correlation function (cf.\ figure~\ref{fig:autocorr_D}) of a time 
dependent variable $\phi(t)$ is defined as
\begin{equation}\label{eq:autocorr}
R_{\phi\phi}\left(\tau\right) = \frac{\int_{a}^{b}\phi''(t)\phi''(t+\tau)\wrt{}t}%
{\int_{a}^{b} \phi''(t)\phi''(t) \wrt{}t}.
\end{equation}
It can be seen that the auto-correlation of both vertical and horizontal 
velocities of the spheroid with $\gls{chi}=1.1$ decay rapidly on a time scale of 
${\cal O} (10)$ gravitational units, and then maintain a periodic behavior with 
small amplitude.
Similar behavior is found for the vertical component of the angular velocity, whereas
the horizontal component exhibits a more intense oscillation periodic correlation
correlated over longer times.
For the larger aspect ratio ($\gls{chi}=1.5$) the auto-correlation functions for the
linear velocities show a clear decay and lack of periodicity.
Both angular velocity components of the case with $\gls{chi}=1.5$ show a damped periodic 
behavior similar to the horizontal angular velocity of the case with $\gls{chi}=1.1$.

\section{Extension of the immersed boundary approach for non-spherical particles}
\label{sec:ibm/method}

Here we present an extension of the direct forcing \gls{ibm} proposed by 
\cite{uhlmann:2005} in order to simulate the presence of non-spherical 
particles in an incompressible flow.
In this formulation the non-conforming feature of the grid with respect to the
particle is treated with a direct-forcing approach, in which a forcing term 
\gls{vfibm} is included on the right hand side of the momentum equation 
\eqref{eq:gov_mom} to model the presence of the particle.
Time marching is performed with a three-step Runge-Kutta scheme for the 
advective term and a Crank-Nicholson scheme for the viscous term.
Continuity is enforced by using a fractional-step method, and spatial 
discretization is done by finite-differences of second order on a uniform, 
staggered grid.
The interpolation/spreading steps between the Eulerian (fluid) and Lagrangian
(particle) grids make use of a regularized Dirac delta function 
\citep{peskin:1972}; in particular we use the variant with a support of three
grid points as proposed by \cite{roma:1999}.
%

In its original description \citep{uhlmann:2005}, the algorithm takes advantage
of the isotropic shape of spherical particles in two different aspects:
\begin{itemize}
\item Rotations need not to be tracked due to the isotropy of a spherical particle.
\item The angular momentum $\gls{I}\gls{vomep}$ is parallel to the angular
      velocity \gls{vomep}, so the integration of the angular momentum equation
     \eqref{eq:motion_rot} is simplified.
\end{itemize}
Both assumptions become invalid when dealing with non-spherical particles; 
therefore, here we propose an extension of the algorithm.
Since the treatment of the fluid part (equation \ref{eq:gov}) as well as the linear 
momentum equation for the particles (equation \ref{eq:motion_lin}) are 
unchanged compared to the original description, they are not included here.
However, the entire algorithm is reproduced in the present \ref{app:ibm} for completeness.

\review{}{%
  Please note that we have previously used essentially the
  same extension in a different context
  and in a different code \citep{arranz:2018},
  where, however, 
  little numerical details
  and only limited evidence of validation 
  were given.}

The first modification of the original algorithm concerns the integration of the 
angular momentum equation \ref{eq:motion_rot} in a body-fixed reference frame, 
in which the moment of inertia tensor \gls{I0} is constant in time such that we 
can write:
\begin{equation}\label{eq:hgi}
\gls{I0}\ddt{\gls{vomep_0}}+\gls{vomep_0}\times\gls{I0}\gls{vomep_0}=
 \gls{R} \left(\gls{rhof}\int \gls{rs}\times \left(\gls{tau}\cdot\vec{n}
                            -\gls{p}\vec{n}\right)\wrt A\right),
\end{equation}
where \gls{R} is the rotation matrix (cf.\ \ref{app:ibm} for more details) and
the subscript $b$ is used to identify quantities expressed in the body-fixed
reference frame.
Assuming full rigid motion of the fluid inside the particle, the discretized
equation for the $k$th substep of the Runge-Kutta scheme used to update the 
angular velocity reads\footnote{This replaces equation (15) in 
\cite{uhlmann:2005}.}:
\begin{equation}\label{eq:ome_disc}
\frac{\gls{vomep_0_k}-\gls{vomep_0_kk}}{\gls{dt}} = %
-\frac{\gls{rhof}}{\gls{rhop}-\gls{rhof}}\gls{rhop}\,\gls{I0inv} \gls{lagT_0_k}
-\gamma_k \gls{I0inv}\left( \gls{vomep_0_kk}\times \gls{I0}\gls{vomep_0_kk} \right)
-\zeta_k    \gls{I0inv}\left( \gls{vomep_0_kkk}\times \gls{I0}\gls{vomep_0_kkk} \right), 
\end{equation}
where \gls{dt} is the time step, \gls{lagT_0} is the contribution to the torque
due to the forcing \gls{vfibm} expressed in the body-fixed coordinate system
(cf.\ \ref{app:ibm} for precise definitions).
The coefficients $\gamma_k$ and $\zeta_k$ are coefficients of the Runge-Kutta
scheme used in \cite{uhlmann:2005}, originally taken from \cite{rai:1991}.
It should be mentioned that if one chooses the orientation of the body-fixed 
reference frame so that it is aligned with the principal axes of the body, the 
inertia tensor and its inverse become diagonal, and the implementation of 
\eqref{eq:ome_disc} is accordingly simplified: the inertia tensor \gls{I0} and 
its inverse \gls{I0inv} can be stored in a buffer of only three positions, the 
components of \gls{I0inv} are simply computed as 
\review{$\gls{I0inv}(i,j)=1/\gls{I0}(i,j)$ and}{%
        $\gls{I0inv}(i,i)=1/\gls{I0}(i,i)$ 
(note that $\gls{I0inv}(i,j)=\gls{I0}(i,j)=0$ for $i\ne j$ when the orientation 
of the body-fixed reference frame is aligned with the principal axes of the body).
Furthermore, }
every operation involving any of \gls{I0} or \gls{I0inv} 
and a vector is computed as the component by component multiplication of two 
vectors.
Also note that the choice of a fully explicit discretization in
\eqref{eq:ome_disc} does not lead to any adverse numerical stability properties.

The next modification of the algorithm is the tracking of the particle rotations.
Here we use a formulation based on quaternions, $\gls{quat}=\left(\gls{q1},%
\gls{q2},\gls{q3},\gls{q4}\right)$, whose components are defined in terms of the
vector defining the rotating axis, \gls{qvec}, and the rotated angle, \gls{qphi},
as $\gls{qi}=\gls{qvec_i}\sin\left(\gls{qphi}/2\right)$ for $i=1,2,3$ and 
$\gls{q4}=\cos\left( \gls{qphi}/2\right)$.
The evolution equation for the quaternion reads \citep{tewari:2007}
\begin{equation}\label{eq:quattime}
\ddt{\gls{quat}} = \frac{1}{2} \gls{QQuat} \gls{quat},
\end{equation}
where the matrix \gls{QQuat} is defined in terms of the angular velocity as
\begin{equation}\label{eq:quatMatrix}
\gls{QQuat} = \left(\begin{array}{cccc}
                                        0 & \hphantom{\shortminus{}}\gls{vomep_0_z} &           \shortminus{} \gls{vomep_0_y} &  \gls{vomep_0_x} \\
             \shortminus{}\gls{vomep_0_z} & \hphantom{\shortminus{}}              0 & \hphantom{\shortminus{}}\gls{vomep_0_x} &  \gls{vomep_0_y} \\
  \hphantom{\shortminus{}}\gls{vomep_0_y} &            \shortminus{}\gls{vomep_0_x} & \hphantom{\shortminus{}}             0  &  \gls{vomep_0_z} \\
             \shortminus{}\gls{vomep_0_x} &            \shortminus{}\gls{vomep_0_y} &           \shortminus{} \gls{vomep_0_z} &             0 
\end{array}\right).
\end{equation}
%
Using the same temporal discretization as in \eqref{eq:ome_disc}, the $k$th 
substage of the Runge-Kutta scheme for the quaternion reads
\begin{equation}
\frac{\gls{quat_k}-\gls{quat_kk}}{\gls{dt}}=\gamma_k\frac{1}{2}\gls{QQuat_kk}\gls{quat_kk}%
                                            +\zeta_k\frac{1}{2}\gls{QQuat_kkk}\gls{quat_kkk}.
\end{equation}

Finally, the distribution of points on the surface of the spheroid has to be 
done considering that the non-spherical particles are not isotropic.
In the case of spheroids one can use a similar approach as the method (2) 
described in appendix A.2.1 of \cite{uhlmann:2005}, but considering geodesic 
distances (along the shortest path restricted to the spheroid's surface) to 
calculate the mutual repulsive energy instead of Euclidean distances; this is
the approach which we have chosen herein.
Details of the procedure can be found in the present
\review{Appendix B}{\ref{app:lagmesh/surface}.} 

\review{}{%
Additionally, for one of the cases discussed below (cf.\
\S~\ref{sec:ibm/results/B}) we employ a distribution of points on the
surface of the spheroid as well as throughout its interior (cf.\
\ref{app:lagmesh/volume}) 
in order to impose the constraint of rigid body motion more strongly,
and thereby to reduce residual flow inside the particle.
It should be mentioned that by using this approach the number of
Lagrangian force points per particle varies as $(\gls{dd}/\gls{dx})^{3}$, 
as compared to a power-2 increase for the standard surface-only forcing. 
As a consequence, forcing the entire volume occupied by the particles
becomes computationally expensive when high solid volume fractions are
to be simulated.
Therefore, it would be worthwhile to explore alternatives in a future
study. One possiblity is to analyze the effectiveness of multi-stage
immersed boundary algorithms \citep[such as][]{luo:2007} on the
residual currents in the configuration of \S~\ref{sec:ibm/results/B}
below. 
}

\review{}{%
It should be mentioned that the choice of integrating the angular equation of 
motion in a body-fixed reference frame as well as using a formulation based on 
quaternions can also be found in the work of
\cite{eshghinejadfard:2016} and of 
\cite{tschisgale:2018} who employ immersed boundary algorithms based on
discrete delta functions for the
description of non-spherical particles in lattice Boltzmann and
finite-difference frameworks, respectively.
\cite{yang:15} have likewise used a combination of body-fixed and
inertial reference frames as well as quaternions in an immersed
boundary method; the latter, however, uses a reconstruction technique
instead of a discrete delta function. 
A substantially different choice for the integration of the angular momentum
equation can be found in the direct forcing \gls{ibm} by \cite{ardekani:2016},
where the rotation matrix (instead of quaternions) is advanced in time in 
an inertial reference frame in which the inertia tensor of the particle is 
computed at each time step by an iterative process.
}

\section{Immersed boundary computations}
\label{sec:ibm}

In this section the results presented in \ref{sec:sem/results} are reproduced 
with the direct forcing \gls{ibm} originally proposed by \cite{uhlmann:2005} and
modified according to section \ref{sec:ibm/method}.
Each case will be computed with increasing spatial resolutions in order to 
perform a grid convergence study for the case of settling spheroidal particles.
%

\subsection{Computational set-up}
\label{sec:ibm/setup}

\begin{figure} 
\makebox[\textwidth][c]{ \includegraphics[scale=1.0]{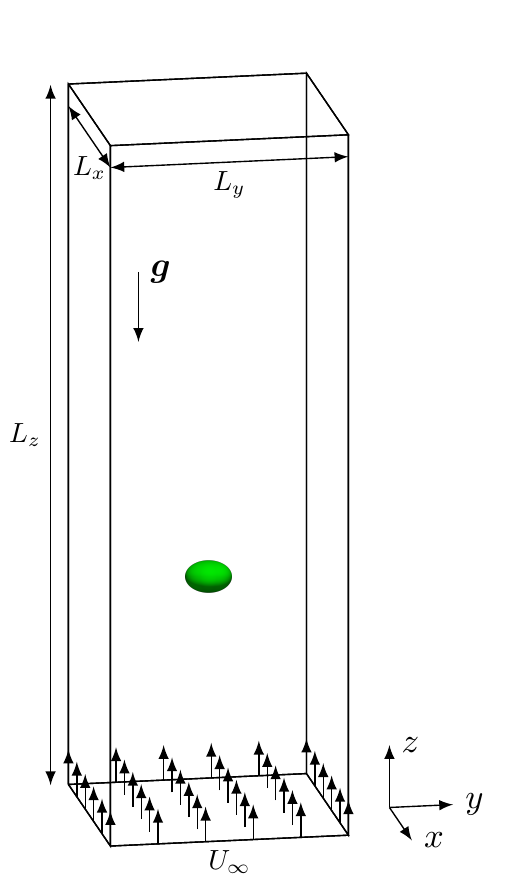} }
\caption{Computational setup for the \gls{ibm} computations with side-lengths
$L_x=L_y=5.33\gls{dd}$ and $L_z=16\gls{dd}$. 
\label{fig:ibm_setup}} 
\end{figure}

The domain is a cuboid of side-lengths $L_x=L_y=5.33\gls{dd}$ and 
$L_z=16\gls{dd}$ as shown in figure \ref{fig:ibm_setup}.
The size of the domain has been selected based on \cite{uhlmann:2014b}.
Regarding the boundary conditions, a uniform vertical velocity $\gls{Uinf}$ is imposed at the 
lower boundary and an advective boundary condition at the top boundary of the 
domain.
Periodic boundary conditions are set in the lateral directions ($x$ and $y$) in
order to allow the particle to freely move in the horizontal directions.

\def\tx{\text{x}}
Each of the cases presented in section \ref{sec:sem/results} is reproduced using
uniform and isotropic spatial resolutions up to $\gls{dd}/\gls{dx}=48$.
This results in a grid size of $[96\,\tx{}\,96\,\tx{}\,288]$ points for the 
lowest resolution and a grid of $[256\,\tx{}\,256\,\tx{}\,768]$ points for the 
highest resolution.
Except where stated otherwise, the time step is set to 
$\gls{dt}\gls{Uinf}/\gls{dd}=\num{1.78e-2}, \num{1.33e-2}, \num{0.89e-2}$ 
and $\num{0.67e-2}$ for the cases with resolution $\gls{dd}/\gls{dx}=18, 24, 36$
and $48$, respectively, which leads to $\gls{CFL}  \lesssim 0.5$.

In order for the particle to remain inside of the computational domain for a
sufficiently long time, the values of \gls{g} and \gls{nu} must be adjusted to match a
given \gls{Ga} and, on average, balance the hydrodynamic force with gravity.
%
A brief description of the steps taken to fulfill these two conditions is:
\begin{enumerate}
   \item Run a simulation in which the particle is fixed and set \gls{nu} to 
         match an (estimated) target \gls{Re}.
   \item From the hydrodynamic force obtained in the previous step, compute
         the value of \gls{g} needed to compensate it as if the particle were 
         free to move. Use this value of \gls{g} to set \gls{nu} to match the 
         desired \gls{Ga} and run a second simulation (particle is still fixed).
   \item Again, compute the value of \gls{g} from the hydrodynamic force obtained in
         the previous step and adjust \gls{nu} to match the desired \gls{Ga}.
         Run a simulation in which the particle is free to move.
         Repeat this step updating the value of gravity depending on the mean
         drift of the particle with respect to the computational domain:
         Reduce (increase) \gls{g} if the particle drifts towards the inlet 
         (outlet).
         In this work, a simple bisection method with a relaxation factor of 
         $0.5$ was successfully employed.
\end{enumerate}

\subsection{Results}
\label{sec:ibm/results}

In the following, errors of converged values are computed for any quantity
$\phi$ 
%
\begin{equation}\label{eq:err}
\varepsilon(\phi) = %
\frac{\left|\phi - \phi_{\text{ref}}\right|}{\phi^\ast_{\text{ref}}},
\end{equation}
%
where the subscript ``ref'' indicates a value taken from the spectral-element
computations presented in section~\ref{sec:sem} \review{}{and the superscript $\ast$ 
is used to indicate that we use $\phi^\ast_\text{ref}=\gls{vup_v}^{\text{ref}}$
for vanishing reference values (e.g.\ those velocities $v$ for which
$v\ll \gls{Ug}$) and $\phi^\ast_\text{ref}=\phi_\text{ref}$ otherwise.
This procedure guarantees that the values obtained for the errors are meaningful
and not amplified.
}
\review{
Definition \eqref{eq:err} is obviously not meanigful for vanishing reference values
$\phi_{\text{ref}}=0$; it is also misleading for quantities which are very small 
compared to some global reference quantity (e.g.\ those velocities $v$ for which
$v\ll \gls{Ug}$). 
}{}
Therefore, we use the vertical reference velocity component $\gls{vup_v}^{\text{ref}}$
in the denominator of \eqref{eq:err} for the normalization of \gls{vup_h} and 
\gls{vup_hzp}, and similarly for \gls{vomep_hzp}
\citep[cf.][p.\ 234, for an extended discussion]{uhlmann:2014b}.

\subsubsection{Steady vertical regime}
\label{sec:ibm/results/A}

\begin{table}
\begin{center}
\caption{
\gls{ibm} results of steady vertical regime for spheroid with $\gls{chi}=1.1$
($\gls{Ga}=100$, $\gls{kappa}=2.10$) for different spatial resolutions and time steps.
Reference values are taken from table \ref{tab:results_refA}.
\label{tab:results_ibmA}}
\begingroup \footnotesize %
\begin {tabular}{|cc|cc|cc|c|}%
\toprule $\gls {dd}/\gls {dx}$&$\gls {CFL}$&\gls {vup_v}&$\varepsilon $&$\gls {Lr}/\gls {dd}$&$\varepsilon $&$\langle \varepsilon (\gls {vur_ll})\varepsilon (\gls {vur_ll})\rangle _z^{1/2}$\\\midrule %
\ensuremath {18}&\ensuremath {0.52}&\ensuremath {-1.6449}&\ensuremath {0.0246}&\ensuremath {1.8492}&\ensuremath {0.0173}&\ensuremath {0.0027}\\%
\ensuremath {24}&\ensuremath {0.52}&\ensuremath {-1.6552}&\ensuremath {0.0184}&\ensuremath {1.8707}&\ensuremath {0.0292}&\ensuremath {0.0019}\\%
\ensuremath {36}&\ensuremath {0.52}&\ensuremath {-1.6661}&\ensuremath {0.0120}&\ensuremath {1.8826}&\ensuremath {0.0357}&\ensuremath {0.0012}\\%
\ensuremath {48}&\ensuremath {0.52}&\ensuremath {-1.6717}&\ensuremath {0.0087}&\ensuremath {1.8848}&\ensuremath {0.0369}&\ensuremath {0.0010}\\%
\midrule \ensuremath {24}&\ensuremath {0.26}&\ensuremath {-1.6653}&\ensuremath {0.0125}&\ensuremath {1.8547}&\ensuremath {0.0204}&\ensuremath {0.0013}\\%
\ensuremath {24}&\ensuremath {0.13}&\ensuremath {-1.6702}&\ensuremath {0.0095}&\ensuremath {1.8456}&\ensuremath {0.0153}&\ensuremath {0.0010}\\%
\midrule \ensuremath {18}&\ensuremath {0.06}&\ensuremath {-1.6604}&\ensuremath {0.0153}&\ensuremath {1.8213}&\ensuremath {0.0020}&\ensuremath {0.0017}\\%
\ensuremath {24}&\ensuremath {0.06}&\ensuremath {-1.6722}&\ensuremath {0.0084}&\ensuremath {1.8420}&\ensuremath {0.0134}&\ensuremath {0.0009}\\%
\ensuremath {36}&\ensuremath {0.06}&\ensuremath {-1.6843}&\ensuremath {0.0012}&\ensuremath {1.8535}&\ensuremath {0.0197}&\ensuremath {0.0004}\\%
\ensuremath {48}&\ensuremath {0.06}&\ensuremath {-1.6906}&\ensuremath {0.0026}&\ensuremath {1.8605}&\ensuremath {0.0235}&\ensuremath {0.0003}\\\bottomrule %
\end {tabular}%
\endgroup %
\end{center}
\end{table}

\begin{figure}
\makebox[\textwidth][c]{
\includegraphics[scale=1.00]{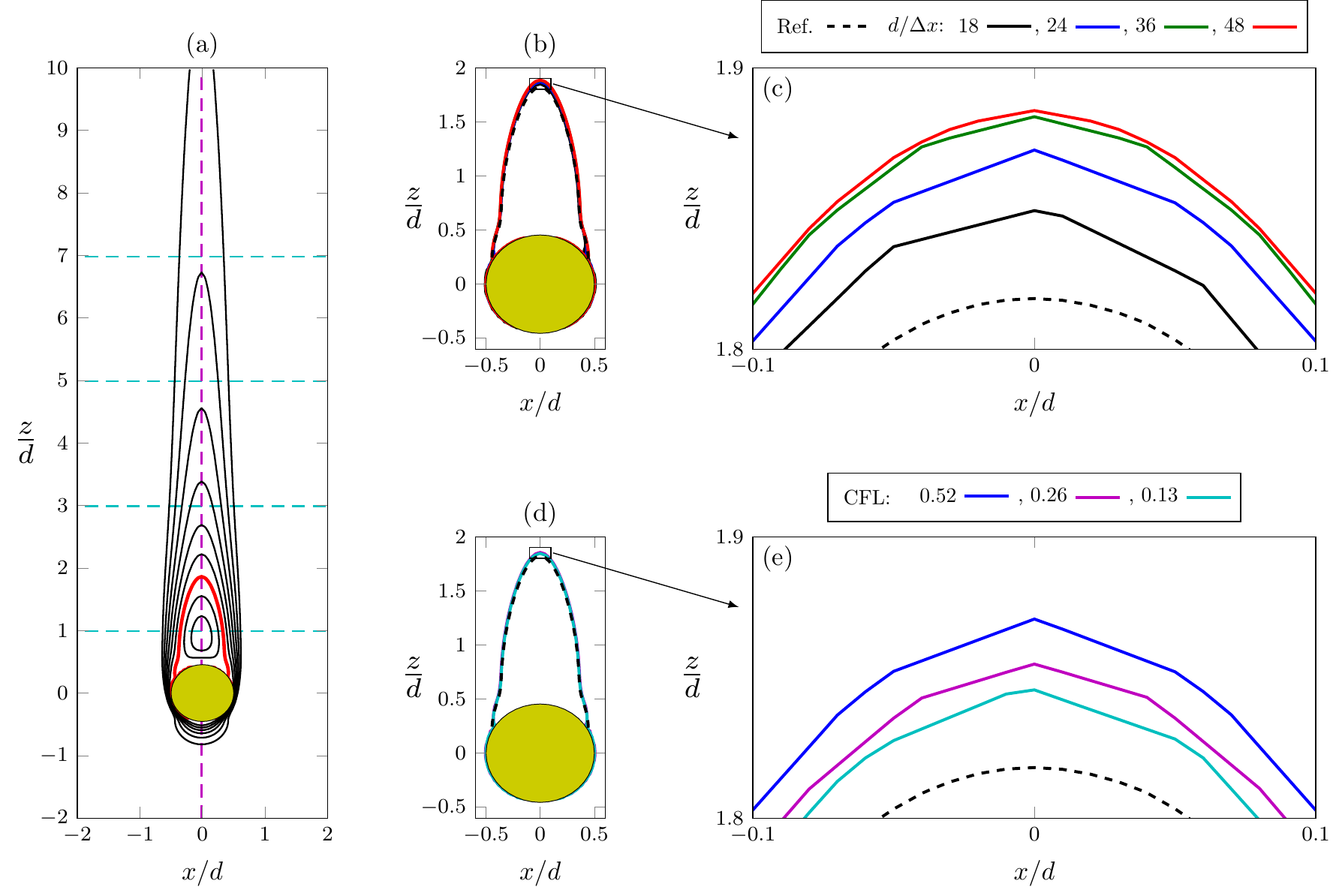}}
\caption{%
a) Iso contours of relative velocity $\gls{vur_ll}$ for the steady vertical case 
\gls{A11M100} ($\gls{Ga}=100$, $\gls{chi}=1.1$ and $\gls{mstar}=1$)
using the \gls{ibm} with a resolution of $\gls{dd}/\gls{dx}=24$.
The contour $\gls{vur_ll}=0$ is highlighted in red.
b) Recirculation region ($\gls{vur_ll}=0$) for different spatial resolutions of the same case and
$\gls{CFL}=0.52$ (zoom inset shown in c).
d) Recirculation region for different time steps with $\gls{dd}/\gls{dx}=24$ (zoom inset shown in e).
\label{fig:visu_SV_ibm}}
\end{figure}

Table \ref{tab:results_ibmA} shows the terminal falling velocity \gls{vup_v} and
the recirculation length \gls{Lr} obtained using different spatial resolutions
and \gls{CFL}.
It can be seen that the error in the terminal velocity \gls{vup_v} is 
consistently reduced towards the reference value,
except for the case with the highest spatial and temporal 
resolution ($\gls{dd}/\gls{dx}=48$, $\gls{CFL}=0.06$),
in which the value of \gls{vup_v} is slightly overestimated.
%
Please note that the error for such refined simulations is already
well below one percent, and that minor differences in the two setups
(e.g.\ the slight variation in the blockage ratio due to different
cross-sections of the computational domain) might become relevant.
%
On the contrary, the recirculation length \gls{Lr} seems to converge to a 
slightly different value than that of the spectral-element solution.
We will return to this point shortly.
Figure \ref{fig:visu_SV_ibm}a shows isocontours of relative velocity 
\gls{vur_ll} for the case with $\gls{dd}/\gls{dx}=24$, highlighting the extent
of the recirculation bubble ($\gls{vur_ll}=0$).
Figures \ref{fig:visu_SV_ibm}b and c show the extent of the recirculation bubble
for all the spatial resolutions.
Note that, in the same fashion as in figures \ref{fig:visu_SV} and 
\ref{fig:visu_SO}, the dashed lines in figure \ref{fig:visu_SV_ibm} represent the
location at which the velocities supplied in the additional material are 
sampled.

It can be seen in figure~\ref{fig:visu_SV_ibm}(c) that under spatial grid
refinement (while keeping the CFL number fixed) the IBM computations converge to
a solution which exhibits a slight discrepancy as compared to the spectral-element 
reference solution.
When decreasing the CFL number, however, the solution for a fixed spatial 
resolution (figure~\ref{fig:visu_SV_ibm}e) can be seen to successively approach
the reference solution.
The convergence is quantified in table~\ref{tab:results_ibmA}, where both errors
in \gls{vup_v} and \gls{Lr} are seen to decrease with decreasing time step.
Note that although the solution is steady in the frame of reference moving with
the particle, it is still unsteady in the frame attached to the grid in the 
\gls{ibm} method. 
The present observation of a necessity to refine the time step in order to 
obtain convergence for very fine grids has already been made previously in the 
context of settling spheres by \cite{uhlmann:2014b}, cf.\ their \S~3.2.2.
More recently, \cite{zhou:2019b} have analyzed the spatio-temporal convergence
of the direct forcing immersed boundary method in more detail.
They have confirmed theoretically and in terms of numerical experiments that
satisfying the \gls{CFL} condition alone does not guarantee the solution to 
converge under spatial grid refinement.
Rather, one needs to make sure that the time step is sufficiently small with
respect to the characteristic time scales of the problem at the fluid-solid
interface.

\begin{figure}
\begin{center}
\makebox[\textwidth][c]{
\includegraphics{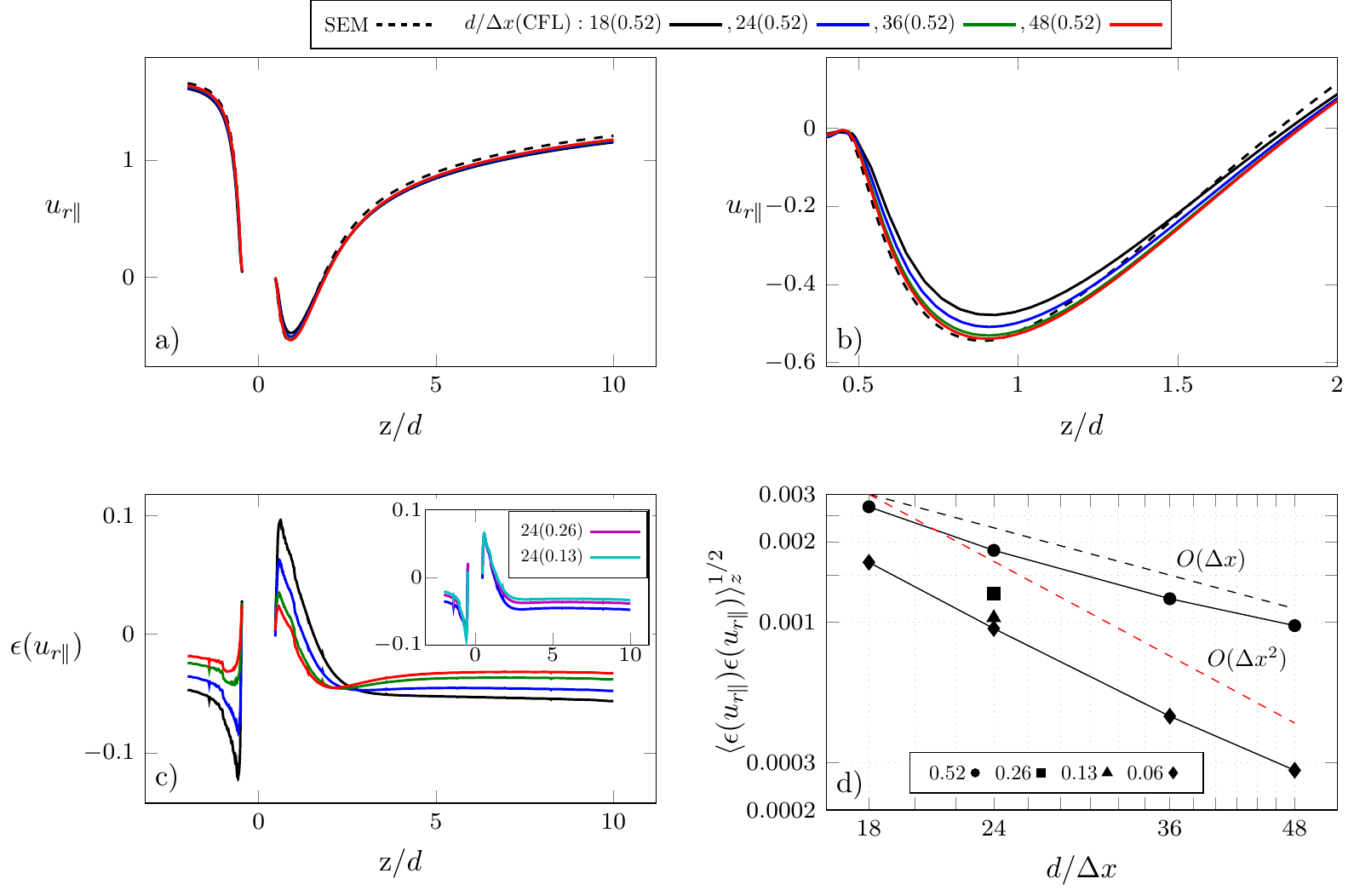}
}
\caption{%
a) Profiles of \gls{vur_ll} for steady vertical regime ($\gls{Ga}=100$, $\gls{chi}=1.1$, 
$\gls{mstar}=1$) on the vertical axis through the particle center.
b) Zoom on the wake next to the particle shown in a).
c) Difference of \gls{ibm} computations with respect to the reference solution
$\varepsilon(\gls{vur_ll})=\gls{vur_ll}^{\text{IBM}}-\gls{vur_ll}^{\text{SEM}}$.
The inset in c) contains the additional cases with resolution $\gls{dd}/\gls{dx}=24$ 
and smaller time step ($\gls{CFL}=0.26,0.13$) together with the case with $\gls{CFL}=0.52$.
d) Convergence of the error shown in panel c averaged over the vertical direction.
The legend in d) indicates the value of \gls{CFL}.
\label{fig:Lr_A_centerline}}
\end{center}
\end{figure}

Despite the previous remarks on convergence in the immediate vicinity of the
immersed solid object, the solution does indeed get in general much better with
increasing spatial resolution, even while keeping the value of the \gls{CFL}
number constant.
This can be seen e.g.\ when considering the relative velocity \gls{vur_ll} 
along the vertical axis passing through the particle center, as shown in 
figure~\ref{fig:Lr_A_centerline}.
In particular the error plotted in panel~\ref{fig:Lr_A_centerline}d clearly 
decreases with increasing spatial resolution.
More specifically, it can be observed that the error changes from 
overpredicting the value of \gls{vur_ll} for $z/\gls{dd}<1$ to underpredicting
it for $z/\gls{dd}>2$.
%
The r.m.s.\ error $\langle\varepsilon(\gls{vur_ll})\varepsilon(\gls{vur_ll})\rangle_z^{1/2}$ 
(evaluated for the intervals $-2\leq z/d \leq -0.5/\gls{chi}$ and 
$0.5/\gls{chi}\leq z/d \leq 10$) is included in table~\ref{tab:results_ibmA}: 
\review{it is seen to decrease roughly linearly with $\Delta x$.}{%
it shows first order convergence with $\Delta x$.}

\begin{figure}
\makebox[\textwidth][c]{\includegraphics{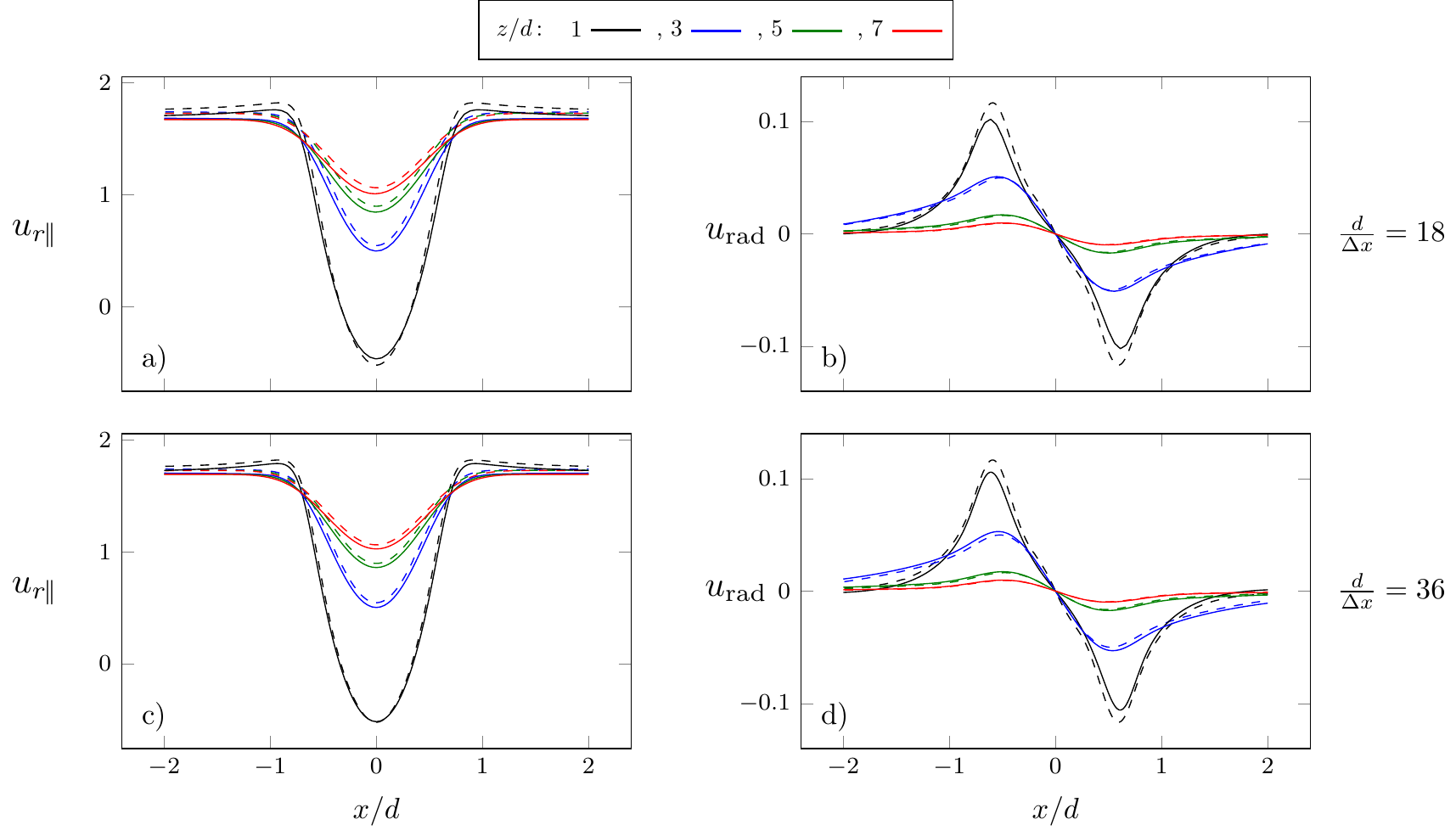}}
\caption{%
Cross profiles of vertical and horizontal components of the relative velocity
for steady vertical regime case \gls{A11M100} ($\gls{Ga}=100$, $\gls{chi}= 1.1$, $\gls{mstar}=1$)
at different positions of $z$ (see legend) 
obtained with the \gls{ibm} with a resolution of $\gls{dd}/\gls{dx}=18$ (a and 
b) and $\gls{dd}/\gls{dx}=36$ (c and d), both with $\gls{CFL}=0.52$.
The reference solution is represented with dashed lines.
\label{fig:profiles_SV_cross_ibm}}
\end{figure}


Velocity profiles of \gls{vur_ll} and \gls{vur_r} along lines perpendicular to 
the vertical direction at different positions of $z$ (as indicated in 
figure \ref{fig:visu_SV_ibm}) are shown in figure \ref{fig:profiles_SV_cross_ibm}
for two different resolutions.
In both cases the qualitative features of the flow are captured and the small
quantitative discrepancies found when $\gls{dd}/\gls{dx}=18$ are significantly
reduced by increasing the resolution.

\begin{figure}
\begin{center}
\includegraphics{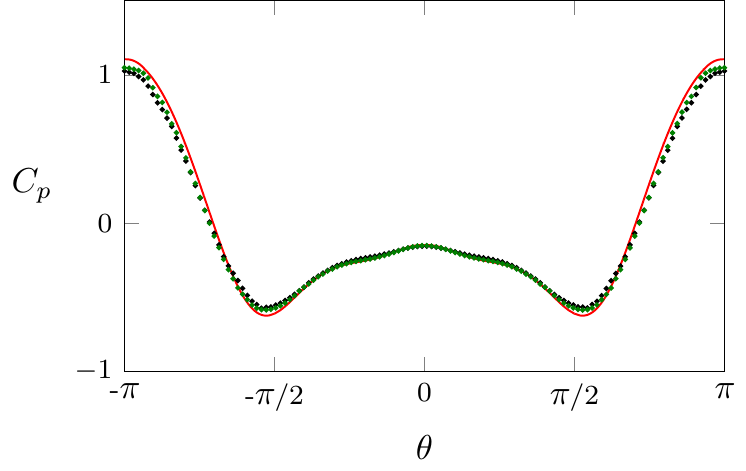}
\caption{%
Pressure coefficient along the great circle for the steady vertical regime
case \gls{A11M100} ($\gls{chi}=1.1$, $\gls{Ga}=100$ and $\gls{mstar}=1$)
 obtained from the \gls{ibm}
computations with resolutions $\gls{dd}/\gls{dx}=18$ (black circles) and 
$36$ (green circles).
Reference solution obtained with \gls{sem} is represented with a solid red 
line.
\label{fig:greatcircle_SV}}
\end{center}
\end{figure}

Finally, the non-dimensional pressure coefficient ${C_p=(p-p_\infty)/(0.5\gls{rhof}\gls{Uinf}^2)}$ 
along the great circle is represented in figure \ref{fig:greatcircle_SV}.
Similarly as in the case of spheres \citep{uhlmann:2014b}, the larger differences
of the case with low resolution ($\gls{dd}/\gls{dx}=18$) are found in the 
vicinity of the stagnation point ($\theta=\pm \pi$).
With double resolution ($\gls{dd}/\gls{dx}=36$), the profile obtained from 
the \gls{ibm} computation is in excellent agreement with the reference 
solution.

\subsubsection{Steady oblique regime}
\label{sec:ibm/results/B}


Table \ref{tab:results_ibmB} shows the \gls{ibm} results of the steady oblique 
regime cases.
In addition to \gls{vup_v} and \gls{Lr}, the values of \gls{vup_h} and 
\gls{alpha_p} are included in the table to characterize the lateral drift of 
the particle as well as the tilting of its symmetry axis.
For the spheroid with $\gls{chi}=1.1$ the values of \gls{vup_h} and 
\gls{alpha_p} obtained with spatial resolutions $\gls{dd}/\gls{dx}=18$ or $24$
are one order of magnitude smaller than the reference value.
When the spatial resolution is increased to $\gls{dd}/\gls{dx}=36$ and $48$.
the agreement with the spectral-element solution is remarkable.
This means that the regime is only marginally captured by the \gls{ibm} using 
spatial resolutions $\gls{dd}/\gls{dx}=18,24$.
For the spheroid with $\gls{chi}=1.5$, the error in \gls{vup_v}, \gls{vup_h} and
\gls{alpha_p} is systematically reduced with increasing spatial resolution.
We observe a similar trend in the value of \gls{Lr} as in the steady vertical cases
(discussed in \S~\ref{sec:ibm/results/A}).

\begin{table}
\begin{center}
\caption{
\gls{ibm} results of the steady oblique regime for spheroids with 
$\gls{chi}=1.1$ ($\gls{Ga}=115$, $\gls{mstar}=1$, $\gls{kappa}=2.1$) and 
$\gls{chi}=1.5$ ($\gls{Ga}=110$, $\gls{mstar}=0.75$, $\gls{kappa}=2.14$).
Reference values are taken from table \ref{tab:results_refB}.
\label{tab:results_ibmB}}
\makebox[\textwidth][c]{
\begingroup \footnotesize %
\begin {tabular}{|cc|cc|cc|cc|cc|}%
\toprule Case&$\gls {dd}/\gls {dx}$&\gls {vup_v}&$\varepsilon $&\gls {vup_h}&$\varepsilon $&$\gls {alpha_p}(^\circ )$&$\varepsilon $&$\gls {Lr}/d$&$\varepsilon $\\\midrule %
\gls {B11M100}&\ensuremath {18}&\ensuremath {-1.7193}&\ensuremath {0.0229}&\ensuremath {0.0041}&\ensuremath {0.0301}&\ensuremath {0.357}&\ensuremath {0.8706}&\ensuremath {1.9878}&\ensuremath {0.0110}\\%
&\ensuremath {24}&\ensuremath {-1.7305}&\ensuremath {0.0165}&\ensuremath {0.0050}&\ensuremath {0.0296}&\ensuremath {0.176}&\ensuremath {0.9361}&\ensuremath {2.0163}&\ensuremath {0.0254}\\%
&\ensuremath {36}&\ensuremath {-1.7368}&\ensuremath {0.0129}&\ensuremath {0.0552}&\ensuremath {0.0011}&\ensuremath {2.636}&\ensuremath {0.0434}&\ensuremath {2.0351}&\ensuremath {0.0350}\\%
&\ensuremath {48}&\ensuremath {-1.7409}&\ensuremath {0.0107}&\ensuremath {0.0641}&\ensuremath {0.0040}&\ensuremath {2.938}&\ensuremath {0.0664}&\ensuremath {2.0385}&\ensuremath {0.0367}\\%
\toprule \gls {B15M075}&\ensuremath {18}&\ensuremath {-1.6514}&\ensuremath {0.0182}&\ensuremath {0.0832}&\ensuremath {0.0177}&\ensuremath {4.090}&\ensuremath {0.2309}&\ensuremath {2.0739}&\ensuremath {0.0171}\\%
&\ensuremath {24}&\ensuremath {-1.6581}&\ensuremath {0.0142}&\ensuremath {0.0975}&\ensuremath {0.0092}&\ensuremath {4.676}&\ensuremath {0.1206}&\ensuremath {2.0965}&\ensuremath {0.0282}\\%
&\ensuremath {36}&\ensuremath {-1.6639}&\ensuremath {0.0107}&\ensuremath {0.1043}&\ensuremath {0.0051}&\ensuremath {4.907}&\ensuremath {0.0773}&\ensuremath {2.1137}&\ensuremath {0.0366}\\%
&\ensuremath {48}&\ensuremath {-1.6681}&\ensuremath {0.0082}&\ensuremath {0.1039}&\ensuremath {0.0054}&\ensuremath {4.890}&\ensuremath {0.0805}&\ensuremath {2.1086}&\ensuremath {0.0341}\\%
\toprule \gls {B15M075} Forc. in.&\ensuremath {24}&\ensuremath {-1.6614}&\ensuremath {0.0122}&\ensuremath {0.1164}&\ensuremath {0.0020}&\ensuremath {5.710}&\ensuremath {0.0737}&\ensuremath {2.0757}&\ensuremath {0.0180}\\%
&\ensuremath {36}&\ensuremath {-1.6773}&\ensuremath {0.0028}&\ensuremath {0.1188}&\ensuremath {0.0035}&\ensuremath {5.628}&\ensuremath {0.0583}&\ensuremath {2.0814}&\ensuremath {0.0207}\\%
&\ensuremath {48}&\ensuremath {-1.6843}&\ensuremath {0.0014}&\ensuremath {0.1198}&\ensuremath {0.0041}&\ensuremath {5.600}&\ensuremath {0.0532}&\ensuremath {2.0832}&\ensuremath {0.0216}\\\bottomrule %
\end {tabular}%
\endgroup %
}
\end{center}
\end{table}

\begin{figure}
\begin{center}
\makebox[\textwidth][c]{
\includegraphics[]{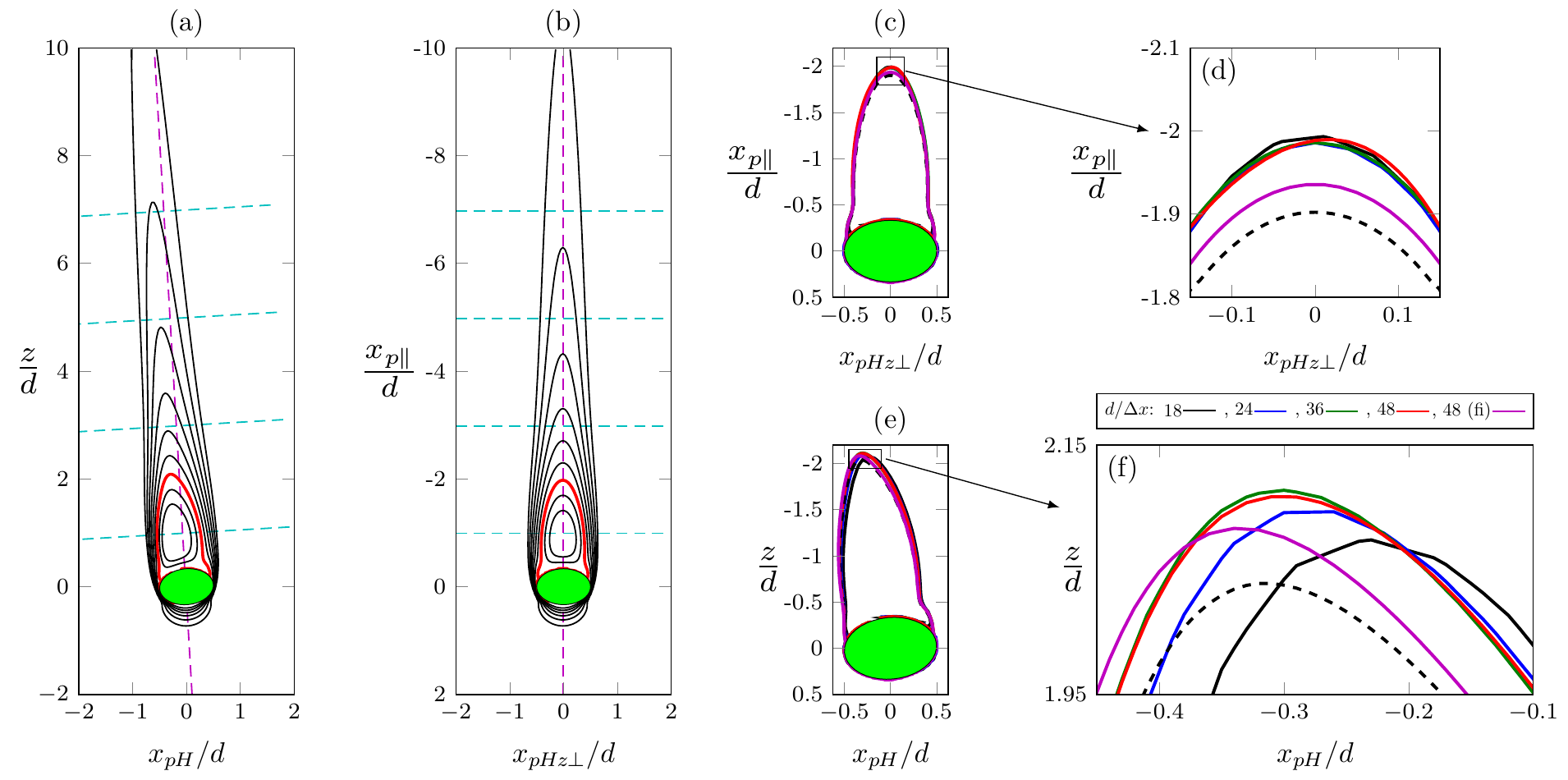}
}
\caption{%
a-b) Iso contours of relative velocity \gls{vur_ll} for the steady 
oblique regime case \gls{B15M075} ($\gls{chi}=1.5$, $\gls{Ga}=110$ and 
$\gls{kappa}=2.14$) with a spatial resolution of $\gls{dd}/\gls{dx}=24$.
The contour $\gls{vur_ll}=0$ is highlighted in red. c,e) Recirculation region
($\gls{vur_ll}=0$) for different spatial resolutions of the same case (zoom
inset shown in d,f).
\label{fig:visu_SO_ibm_bubble}}
\end{center}
\end{figure}

\begin{figure} 
\makebox[\textwidth][c]{ 
\includegraphics[scale=1.0]{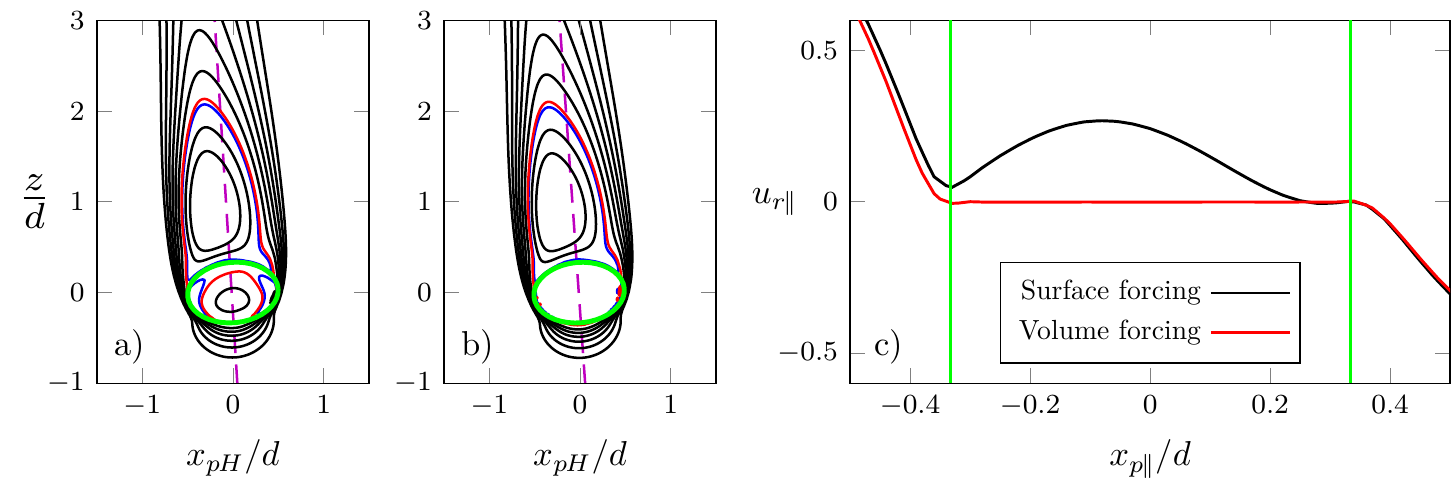}
} 
\caption{
a) Unmasked contours of relative velocity \gls{vur_ll} of steady oblique case
with $\gls{Ga}=110$, $\gls{chi}=1.5$, $\gls{kappa}=2.14$ with spatial resolution 
$\gls{dd}/\gls{dx}=36$ 
distributing Lagrangian force points on the particle surface only,
and b) forcing throughout the volume occupied by the solid particle.
c) Profile of relative velocity \gls{vur_ll} along a line parallel to the 
trajectory passing through the center of the particle ($\gls{xpp}=\gls{xpHzp}=0$)
for both cases, represented with a dashed line in a,b. 
The black contours shown in a,b correspond to $\gls{vur_ll}=-0.4,-0.2,0.2:0.2:1.2$ 
and the blue and red contours to $\gls{vur_ll}=-0.02$  and $0.02$, respectively.
The green line in the three panels represents the surface of the spheroid.
\label{fig:detail_ForcingInside_r36B}} 
\end{figure}

Figure \ref{fig:visu_SO_ibm_bubble} shows isolines of constant velocity \gls{vur_ll}
for the steady oblique regime cases of the spheroid with $\gls{chi}=1.5$.
Panels \ref{fig:visu_SO_ibm_bubble}a-b show contours for the case with spatial resolution 
$\gls{dd}/\gls{dx}=24$ highlighting the recirculation bubble ($\gls{vur_ll}=0$).
Panels \ref{fig:visu_SO_ibm_bubble}c-f show the recirculation bubble obtained for 
different spatial resolutions and an additional case in which forcing points are
distributed also in the interior of the particle (cf.\ \ref{app:lagmesh}).
The motivation to carry out this additional case is that we observe a small
amplitude path oscillation for the case with the highest spatial resolution 
($\gls{dd}/\gls{dx}=48$), 
which can be appreciated in the small lateral 
deviation of the recirculation bubble shown in figure  \ref{fig:visu_SO_ibm_bubble}d.
%
%
\review{It can be seen how the shape of the recirculation bubble is notably better 
captured when forcing inside.}{%
Inspection of the volume occupied by the particle (cf.\ figures~\ref{fig:detail_ForcingInside_r36B}$a$ and $c$) 
reveals that some parasitic flow develops therein.
We have therefore recomputed this case while applying the rigid-body forcing 
throughout the volume occupied by the particle.
As can be seen from figures~\ref{fig:detail_ForcingInside_r36B}$(b,c)$, the parasitic 
flow is effectively suppressed. 
As a consequence it can be seen in figure~\ref{fig:visu_SO_ibm_bubble} how the shape of the 
recirculation bubble is notably better captured when forcing inside.
}
Cases with forcing in the interior of the particle with spatial resolutions
$\gls{dd}/\gls{dx}=24, 36$ are further discussed. 

\begin{figure}
\begin{center}
\makebox[\textwidth][c]{
\includegraphics[scale=0.9]{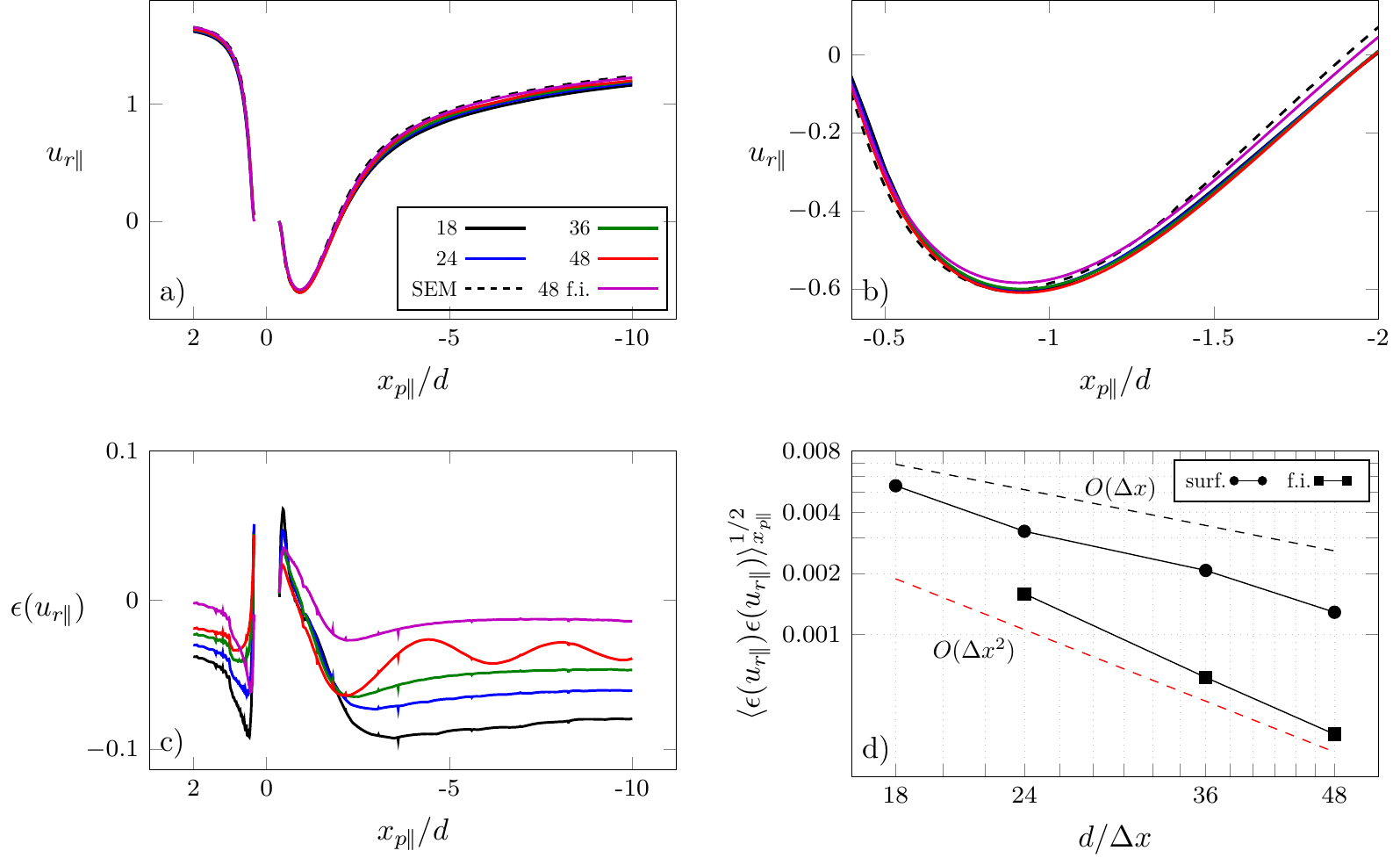}
}
\caption{%
a) Profiles of \gls{vur_ll} for steady oblique regime ($\gls{Ga}=110$, $\gls{chi}=1.5$, 
$\gls{kappa}=2.14$) along a line parallel to the trajectory passing through the 
center of the particle ($\gls{xpp}=\gls{xpHzp}=0$).
b) Zoom of the wake shown in panel a.
c) Difference of \gls{ibm} computations with respect to the reference solution
$\varepsilon(\gls{vur_ll})=\gls{vur_ll}^{\text{IBM}}-\gls{vur_ll}^{\text{SEM}}$.
d) Convergence of the error shown in panel (c) averaged over the line parallel
to the trajectory passing throught the particle's center.
The legend in (d) indicates the forcing point distribution are distributed on the surface (``surf'')
or in the entire volume of the spheroid (``f.i.'').
\label{fig:Lr_B_centerline}}
\end{center}
\end{figure}

Figure \ref{fig:Lr_B_centerline} shows the relative velocity \gls{vur_ll} on 
a line parallel to the trajectory of the particle ($\gls{epll}$) passing through
the center of the particle for all spatial resolutions employed in the steady
oblique case with $\gls{chi}=1.5$.
The error $\varepsilon(\gls{vur_ll})$ is represented in figure 
\ref{fig:Lr_B_centerline}c where, again, we confirm that 
$\langle\varepsilon(\gls{vur_ll}) \varepsilon(\gls{vur_ll}) \rangle_z^{1/2}$ 
systematically decreases with increasing spatial resolution (at constant 
\gls{CFL}).
Figure \ref{fig:Lr_B_centerline}d shows that the convergence of 
$\langle\varepsilon(\gls{vur_ll}) \varepsilon(\gls{vur_ll}) \rangle_z^{1/2}$ 
with the spatial resolution is \review{roughly linear}{of first order}.
The figure also includes the additional cases in which forcing points are 
distributed also in the interior of the particle for spatial resolutions
$\gls{dd}/\gls{dx}=24,36,48$.
The above mentioned oscillations for the highest spatial resolution can be 
observed in figure \ref{fig:Lr_B_centerline}c as well as their absence by 
forcing in the interior of the particle.
Furthermore, figure \ref{fig:Lr_B_centerline}d shows how the convergence of the
error with the spatial resolution approaches second order when forcing also
inside the particle's volume. 

Based on the results obtained for the steady vertical regime and on analogous
work for spheres \citep{uhlmann:2014b}, the higher resolution needed in the case
of the spheroid with lower aspect ratio ($\gls{chi}=1.1$) compared to the higher
aspect ratio ($\gls{chi}=1.5$) is somehow unexpected.
However, a look at the two-parameter state diagram in figure 
\ref{fig:regimesdetail} reveals that the width of the region in which the
steady-oblique regime occurs for a spheroid of $\gls{chi}=1.1$ and 
$\gls{kappa}=2.1$ is of the order of a few units of $\gls{Ga}$, which is
narrower than that of the spheroid with $\gls{chi}=1.5$ and $\gls{kappa}=2.14$.
Furthermore, the selected value of $\gls{Ga}=115$ for the spheroid with 
$\gls{chi}=1.1$ is clearly closer to the critical \gls{Ga} leading to the first
bifurcation (boundary between white and gray regions in figure 
\ref{fig:regimesdetail}a) than the corresponding case with $\gls{chi}=1.5$.
Please recall that small numerical inaccuracies can induce an upward
shift of the threshold leading to the first bifurcation by several Galileo 
number units \citep{uhlmann:2014b}.
Therefore, the higher spatial resolution needed in this case seems reasonable.

\subsubsection{Vertical periodic regime}
\label{sec:ibm/results/C}

Here we evaluate the convergence with spatial resolution of the solution obtained 
by means of the \gls{ibm} for the vertical periodic cases that appear for a
spheroid of aspect ratio $\gls{chi}=1.5$, $\gls{Ga}=150$ and $\gls{kappa}=2.14$.
Table \ref{tab:results_ibmC} shows the statistical moments, as well as errors 
which are computed with respect to the \gls{sem} reference data presented in 
\S~\ref{sec:sem/results/C} (cf.\ statistics in table~\ref{tab:results_refC}).
Note that \gls{vup_hzp} was not included in the reference results presented in table 
\ref{tab:results_refC} because it is zero (the flow maintains planar symmetry).

\begin{figure} 
\makebox[\textwidth][c]{ 
\includegraphics[scale=1.0]{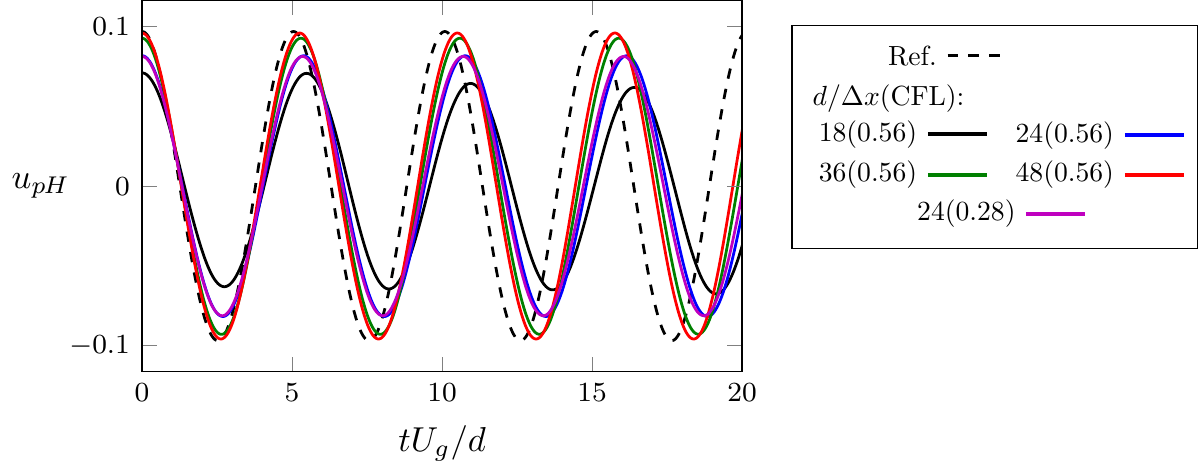} 
}
\caption{Time history of the horizontal particle velocity (\gls{vup_h})
of the vertical periodic \gls{ibm} cases for the spheroid with $\gls{chi}=1.5$,
$\gls{Ga}=150$ and $\gls{kappa}=2.14$ ($\gls{mstar}=0.75$).
\label{fig:timehistory_OSCILLATORY_ibm}} 
\end{figure}

It can be seen that the results generally converge under grid refinement while 
keeping the \gls{CFL} number constant.
However, concerning the average settling velocity \gls{vup_v}, we observe once 
more a convergence to a value which is approximately $1.5\%$ off the reference value (for $\gls{dd}/\gls{dx}=48$) under purely spatial refinement.
Again, when further reducing the time step (i.e. for $\gls{dd}/\gls{dx}=24$ and
reducing \gls{CFL} from  $0.56$ to $0.28$) the solution does approach the reference 
value, as already discussed above (cf.\ \S~\ref{sec:ibm/results/A}).
Figure~\ref{fig:timehistory_OSCILLATORY_ibm} shows the time-evolution of the 
horizontal particle velocity \gls{vup_h}, which is dominated by a single harmonic. 
It can be seen how both the amplitude as well as the period improve under spatial and temporal 
refinement.

\begin{table} 
\caption{%
Results and corresponding error measures of particle-related quantities
for the vertical periodic cases computed by means of the \gls{ibm} with 
$\gls{chi}=1.5$, $\gls{Ga}=150$ and $\gls{mstar}=0.75$ ($\gls{kappa}=2.14$).
\label{tab:results_ibmC}} 
\makebox[\textwidth][c]{ 
\begingroup \footnotesize %
\begin {tabular}{|p{3mm}p{5mm}|p{6mm}p{7mm}|p{10mm}p{7mm}|p{6mm}p{7mm}|p{6mm}p{7mm}|p{6mm}p{7mm}|p{6mm}p{7mm}|p{6mm}p{7mm}|}%
\toprule $\frac {\gls {dd}}{\gls {dx}}$&\gls {CFL}&\gls {St}&$\varepsilon $&\gls {vup_v_m}&$\varepsilon $&\gls {vup_v_a}&$\varepsilon $&\gls {vup_h_a}&$\varepsilon $&\gls {vup_hzp_a}&$\varepsilon $&\gls {vomep_hzp_a}&$\varepsilon $&$\gls {alpha_p_max_deg}$&$\varepsilon $\\\midrule %
\ensuremath {18}&\ensuremath {0.56}&\ensuremath {0.1829}&\ensuremath {0.0782}&\ensuremath {-1.7454}&\ensuremath {0.0030}&\ensuremath {0.0076}&\ensuremath {0.0022}&\ensuremath {0.1434}&\ensuremath {0.0290}&\ensuremath {0.0091}&\ensuremath {0.0053}&\ensuremath {0.2736}&\ensuremath {0.0771}&\ensuremath {6.9324}&\ensuremath {0.2550}\\%
\ensuremath {24}&\ensuremath {0.56}&\ensuremath {0.1857}&\ensuremath {0.0639}&\ensuremath {-1.7307}&\ensuremath {0.0054}&\ensuremath {0.0027}&\ensuremath {0.0006}&\ensuremath {0.1635}&\ensuremath {0.0174}&\ensuremath {0.0004}&\ensuremath {0.0003}&\ensuremath {0.3176}&\ensuremath {0.0518}&\ensuremath {7.7616}&\ensuremath {0.1659}\\%
\ensuremath {36}&\ensuremath {0.56}&\ensuremath {0.1888}&\ensuremath {0.0485}&\ensuremath {-1.7170}&\ensuremath {0.0133}&\ensuremath {0.0039}&\ensuremath {0.0000}&\ensuremath {0.1858}&\ensuremath {0.0046}&\ensuremath {0.0014}&\ensuremath {0.0008}&\ensuremath {0.3479}&\ensuremath {0.0344}&\ensuremath {8.3621}&\ensuremath {0.1014}\\%
\ensuremath {48}&\ensuremath {0.56}&\ensuremath {0.1904}&\ensuremath {0.0405}&\ensuremath {-1.7126}&\ensuremath {0.0158}&\ensuremath {0.0041}&\ensuremath {0.0002}&\ensuremath {0.1919}&\ensuremath {0.0011}&\ensuremath {0.0007}&\ensuremath {0.0004}&\ensuremath {0.3531}&\ensuremath {0.0315}&\ensuremath {8.4161}&\ensuremath {0.0956}\\\hline %
\ensuremath {24}&\ensuremath {0.28}&\ensuremath {0.1869}&\ensuremath {0.0581}&\ensuremath {-1.7493}&\ensuremath {0.0053}&\ensuremath {0.0026}&\ensuremath {0.0007}&\ensuremath {0.1626}&\ensuremath {0.0180}&\ensuremath {0.0005}&\ensuremath {0.0003}&\ensuremath {0.3117}&\ensuremath {0.0552}&\ensuremath {7.5736}&\ensuremath {0.1861}\\\bottomrule %
\end {tabular}%
\endgroup %
}
\end{table} 

\subsubsection{Chaotic regime}
\label{sec:ibm/results/D}

The chaotic regime cases presented in \S~\ref{sec:sem/results/D} are reproduced 
here with a resolution of $\gls{dd}/\gls{dx}=36$.
Table \ref{tab:results_ibmD} shows the statistics of the linear and angular 
particle velocities of the cases with $\gls{chi}=1.1$ ($\gls{Ga}=200$, 
$\gls{mstar}=1$) and $\gls{chi}=1.5$ ($\gls{Ga}=220$, $\gls{mstar}=5$).
For the case with $\gls{chi}=1.1$ the error in the mean vertical velocity is
$1\%$ and for the case with $\gls{chi}=1.5$ as low as $0.2\%$.

\begin{figure}
\begin{center}
\includegraphics{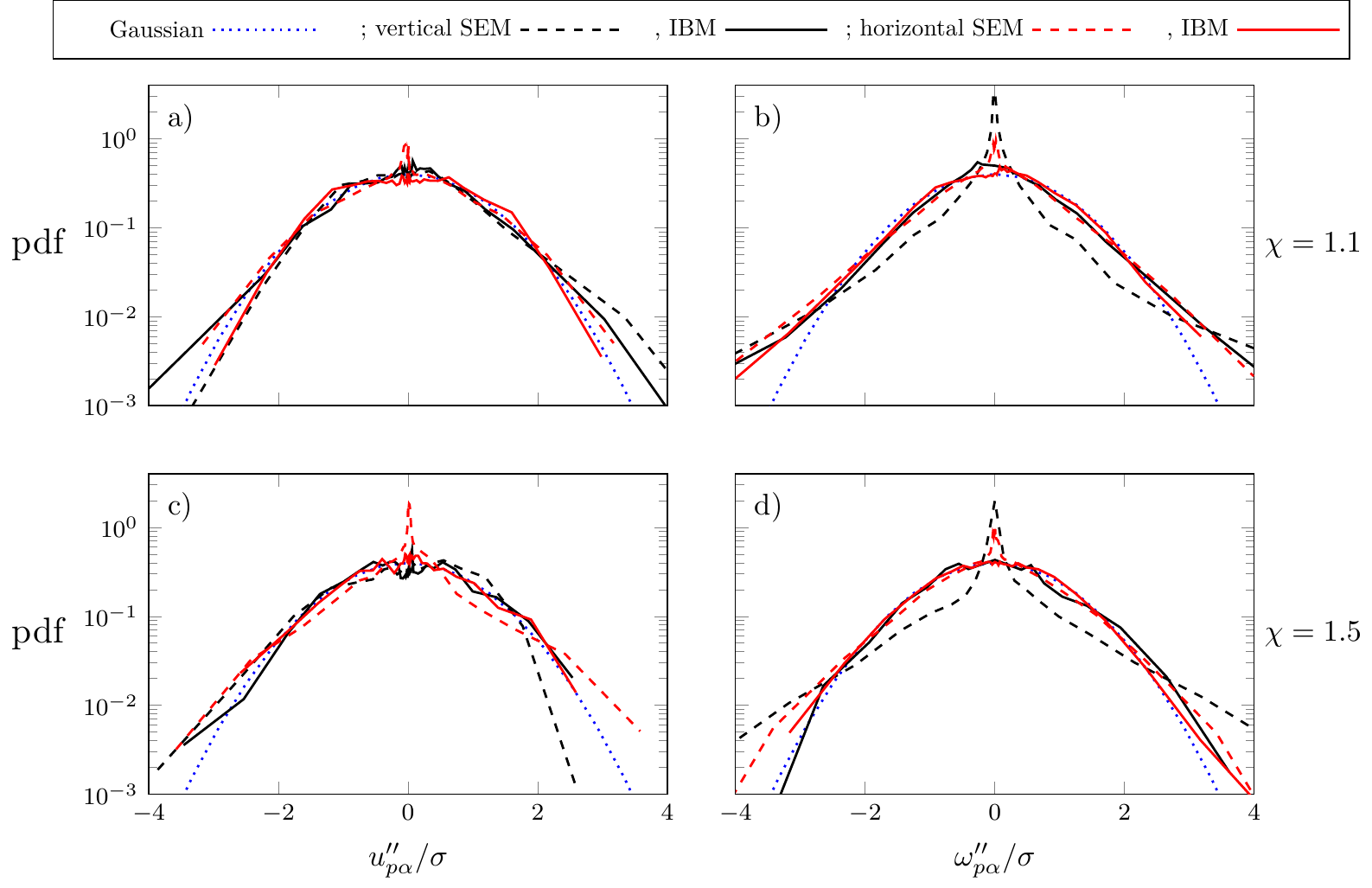}
\caption{%
\gls{ibm} results of chaotic regime for cases \gls{D11M100} (top row) and \gls{D15M500} (bottom row).
\gls{pdf} of vertical and horizontal components of the linear (left panels) and angular (right panels) velocity
are represented together with a Gaussian distribution (see legend).
\label{fig:chaotic_normpdf_u}}
\end{center}
\end{figure}

Figure \ref{fig:chaotic_normpdf_u} shows the \gls{pdf} for both linear and 
angular velocities of the chaotic regime cases computed with the \gls{ibm} 
together with the spectral-element reference solutions.
For the lower aspect ratio ($\gls{chi}=1.1$) the \gls{ibm} computation captures the
almost normal distribution found for the horizontal component of the velocity
as well as the long positive tail in the vertical component and its positive skewness,
whereas the decay of the negative tail is clearly slower compared to the \gls{sem} solution.
The moderate excess kurtosis observed in the horizontal and vertical components
of the angular velocity is also captured.
For the higher aspect ratio ($\gls{chi}=1.5$) the slow decay of the negative 
tail in the pdf of the vertical velocity is captured, but not the fast decay of
the positive tail of the curve.
Regarding the horizontal component of the velocity, the positive kurtosis is
only partially captured by the \gls{ibm} solution.
The horizontal component of the angular velocities of the case with $\gls{chi}=1.5$
shows good agreement with the reference data, whereas the moderate excess 
kurtosis found in the vertical component is not captured.

\begin{figure} 
\makebox[\textwidth][c]{ 
\includegraphics[scale=1.0]{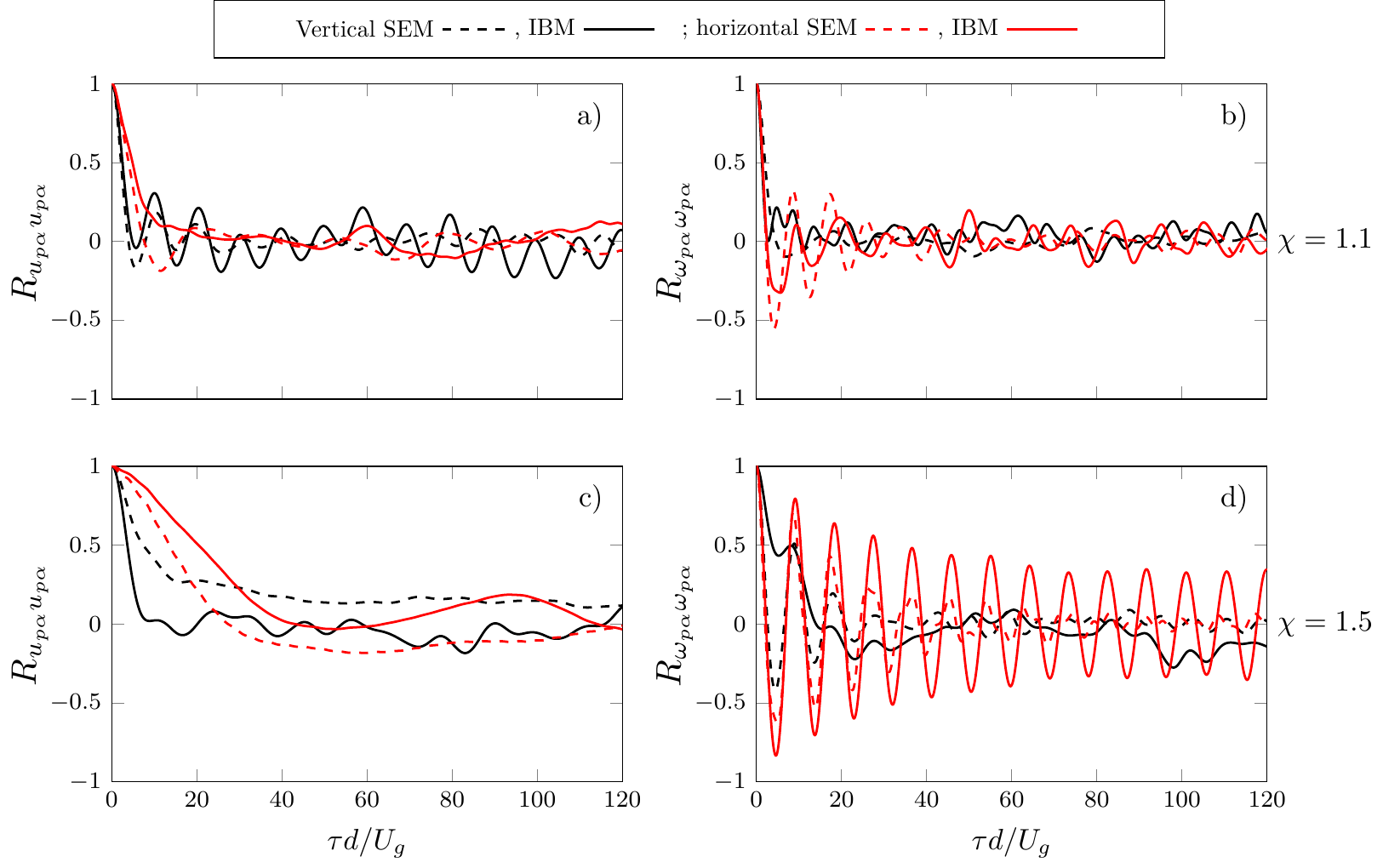} 
}
\caption{%
\gls{ibm} results of chaotic regime for cases \gls{D11M100} (top row) and \gls{D15M500} (bottom row).
Temporal auto-correlations of vertical and horizontal components of the linear (left panels) and angular (right panels) velocity
are represented together with the spectral-element reference data (see legend).
\label{fig:chaotic_autocorr}} 
\end{figure}

Finally, the auto-correlations presented in figure \ref{fig:chaotic_autocorr} 
show a very good agreement for the case with $\gls{chi}=1.1$, where both the 
decay and the frequency of oscillation are well captured.
For the case with $\gls{chi}=1.5$ the decay observed in the vertical velocity
is faster for the \gls{ibm} computation compared to the \gls{sem} solution.
This behavior is switched for the horizontal component.
Also, the slow decay of the vertical component found for separations of 
$\gls{tausep}\gls{Ug}/\gls{dd}\gtrsim 15$ is captured by the \gls{ibm} solution
as an oscillatory behavior.
Regarding the angular velocity of the case with $\gls{chi}=1.5$, the \gls{ibm}
solution features a slower decay than the reference solution in the vertical 
component, although the observed frequency of oscillation is well represented.
The agreement between \gls{ibm} and \gls{sem} in the horizontal component of the
angular velocity of this case is very good; in particular the frequency of the 
oscillation is very well captured.

\begin{table} 
\caption{%
Statistics of linear and angular particle velocities for chaotic cases of spheroids
with $\gls{chi}=1.1$ and $\gls{chi}=1.5$ at $\gls{Ga}=200$ and $\gls{Ga}=220$, respectively,
obtained from immersed-boundary computations with $\gls{dd}/\gls{dx}=36$.
The reference value used to compute the errors is taken from the spectral-element 
solution presented in table \ref{tab:results_refD}.
\label{tab:results_ibmD}} 
\makebox[\textwidth][c]{ 
\begingroup \footnotesize %
\begin {tabular}{|c|cp{7mm}|cp{7mm}|cp{7mm}|cp{7mm}|cp{7mm}|c|}%
\toprule $\gls {chi}$&$\frac {\gls {vup_z_m}}{\gls {Ug}}$&$\quad \varepsilon $&$\frac {\gls {vup_z_f}}{\gls {Ug}}$&$\quad \varepsilon $&$\frac {\gls {vup_x_f}}{\gls {Ug}}$&$\quad \varepsilon $&$\frac {\gls {vop_z_f}}{\gls {Ug}/\gls {dd}}$&$\quad \varepsilon $&$\frac {\gls {vop_x_f}}{\gls {Ug}/\gls {dd}}$&$\quad \varepsilon $&$\frac {\gls {Tobs}\gls {Ug}}{\gls {dd}}$\\\midrule %
\ensuremath {1.1}&\ensuremath {-1.9302}&\ensuremath {0.0118}&\ensuremath {0.0153}&\ensuremath {0.0034}&\ensuremath {0.1009}&\ensuremath {0.0074}&\ensuremath {0.0030}&\ensuremath {0.0002}&\ensuremath {0.0272}&\ensuremath {0.0067}&\ensuremath {1{,}430}\\%
\ensuremath {1.5}&\ensuremath {-1.8656}&\ensuremath {0.0014}&\ensuremath {0.0098}&\ensuremath {0.0014}&\ensuremath {0.0456}&\ensuremath {0.0004}&\ensuremath {0.0014}&\ensuremath {0.0005}&\ensuremath {0.0146}&\ensuremath {0.0001}&\ensuremath {1{,}761}\\\bottomrule %
\end {tabular}%
\endgroup %
}
\end{table}

\section{Conclusions}
\label{sec:conclusions}
%
%

We have presented accurate spectral-element data on low and moderate aspect
ratio oblate spheroids settling under gravity in an unbounded fluid computed
in a comparatively small domain in order to facilitate its use in benchmarking
studies.
\review{}{%
  The data-set which is freely available for download includes all
  aspects of the particle motion, as well as fluid velocity profiles
  and pressure maps.  
}
In this context, we present an extension to the immersed boundary method 
proposed by \cite{uhlmann:2005} in order to handle non-spherical particles, 
which is validated against the spectral-element solutions.

The cases analyzed in this work have been carefully selected to cover the most 
relevant features encountered by finite--size heavy spheroids in wake dominated
flows.
First, the regime of steady vertical motion is considered in which the flow 
field is axi--symmetric.
Then, we analyze the flow after the first regular bifurcation in which the 
axial symmetry of the solution is broken, leading to a steady regime with planar
symmetry.
We also analyze a vertical periodic regime which is not present in the case of 
spheres.
The last regime presented is chaotic, requiring a statistical data analysis.

The extension of the \gls{ibm} originally proposed by \cite{uhlmann:2005} 
presented here makes use of quaternions and a combination of inertial and body-fixed
reference systems to track rotations efficiently without presenting numerical 
instabilities.
Please recall that, although in this work we have only presented results for 
spheroidal particles, the validity of the present algorithm is not restricted to
any particular geometric shape.
Furthermore, we have explored the possibility to apply the immersed boundary 
forcing throughout the volume occupied by the solid particles (instead of only
at the surface) which was found to noticeably improve the results for a given 
spatial resolution.


In the last part of the manuscript we evaluate the accuracy of the enhanced 
immersed boundary algorithm.
For upper end of the steady vertical regime ($\gls{Ga}=100$) the error in the 
particle settling velocity is systematically reduced from approximately $3\%$ 
using $\gls{dd}/\gls{dx}=18$ to approximately $1\%$ when $\gls{dd}/\gls{dx}=48$.
We also demonstrate a convergence of $\mathcal{O}(\gls{dx})$ of the error in the
fluid velocity field when keeping the \gls{CFL} number constant and, 
additionally, how the reduction of the time step for a given spatial resolution
further reduces the error.
Please note that, despite the simplicity of the kinematics obtained in this case,
the flow is characterized by a wake whose recirculation bubble extends
approximately two diameters downstream from the rear stagnation point, implying 
that non-trivial phenomena of flow separation must be accurately captured.
Similar accuracy is achieved for the steady oblique regime, except for the spheroid
with $\gls{chi}=1.1$ and spatial resolutions $\gls{dd}/\gls{dx}=18,24$, where
comparatively low horizontal velocity and particle tilting are obtained.
Here the narrowness of the Ga interval in which this regime appears is making it
hard to be captured with a non-conforming method.
Nevertheless, the loss of axial symmetry and the orientation of the particle indicates
that, despite the quantitative differences, the results are qualitatively in 
agreement with the reference solution.
This highlights the fact that capturing the steady oblique regime (or any regime
that takes place in a narrow region of the two parameter state diagram shown in
figure \ref{fig:regimesdetail}) is already a hard test for a non-conforming-grid
algorithm.
We have also found that IBM forcing on the surface of the particle
leads to an unphysical unsteadiness of the particle motion when using
a very fine grid ($\gls{dd}/\gls{dx}=48$) in a nominally steady case with
relatively small excess particle density ($\chi=1.5$, $Ga=110$, $\tilde{\rho}=2.14$). 
In this case, forcing throughout the volume occupied by the particle 
eliminates the unphysical coupling between the (parasitic) internal flow and
the exterior; it also improves the prediction for a given spatial resolution.
The results obtained for the vertical periodic regime show a systematic 
reduction of the error under spatial refinement, except for the mean settling 
velocity, which exhibits an error of approximately $1.5\%$ when $\gls{dd}/\gls{dx}=24$
is used.
The latter error, however, can be substantially decreased by reducing the time step
at a fixed spatial resolution.
Finally, statistical errors obtained in the chaotic regime using a resolution of
$\gls{dx}/\gls{dd}=36$ are shown to be very small (first and second moment of the
particle velocity are predicted with errors below $1\%$) and probability density
functions as well as auto-correlation functions are reasonably captured.

\review{It is expected that the present study serves as a
  reference benchmarking process for numerical tools aimed to simulate
  flows involving non--spherical particles with significant relative
  velocities. New algorithms can be validated in this fashion, and
  their performance (i.e.\ resolving efficiency) can be gauged.}{%
  It is expected that the present study serves as an additional
  benchmark for numerical tools aimed to simulate flows involving
  non--spherical particles at Reynolds numbers of ${\cal O}(100)$.   
  New algorithms can include the reference cases presented here in
  their benchmarking process in order to gauge their resolving
  efficiency, and hence their performance. 
} 
%
The data-set further allows to precisely determine the required resolution
depending on the tolerance and on the parameter regime of interest.
Last but not least the data-set includes additional parameter points and 
additional physical quantities which have not been published up to this point.
It can therefore be of interest in future work on the collective dynamics of
settling oblate spheroids.
\review{}{%
  Additional potential lies in the analysis of the present simulation 
  results with respect to hydrodynamic force and torque. A number of 
  previous authors have performed parametric studies for fixed
  spheroids placed in uniform inflow, using the simulation data to
  derive empirical force/torque correlation formulas as a function of
  aspect ratio, Reynolds number and relative orientation of the body
  \citep{zastawny:12,ouchene:16,sanjeevi:18,andersson:19}.
  These correlations can then be utilized in Eulerian-Lagrangian
  point-particle simulations beyond the Stokes flow limit
  \citep[e.g.][]{vanwachem:15}. 
  Based on the data for the regime of steady particle motion in the
  present set, it will be possible to enlarge the available parameter
  space. This perspective, as well as further developments applicable
  to the modelling of forces and torque acting on spheroids undergoing
  unsteady motion, should be investigated in future studies. 
}

\review{%
  Note that the present reference data is available under the following URL
  \url{http://www.ifh.kit.edu/dns_data/particles/single_spheroid_sedimentation/}.}{%
  The present reference data is available for download under the following
  persistent DOI: 
  \url{https://dx.doi.org/10.4121/13042793}.
}

\appendix

\section*{Acknowledgments}
The contribution by Manuel Garc\'ia-Villalba to the present
quaternion formulation and its discretization is gratefully acknowledged. 
This work was supported by the German Research Foundation (DFG)
under Project UH 242/11-1.
The computations were partially performed on the supercomputer ForHLR funded by the
Ministry of Science, Research and the Arts Baden--W\"urtemberg and by the
Federal Ministry of Education and Research.

\newpage

\section{Algorithm of the immersed boundary method}
\label{app:ibm}

Each $k$th Runge-Kutta substep of the algorithm ($k=1,2,3$) starts with the computation of the
volume forcing \gls{vfibm} from an explicit estimation of the velocity field (\gls{vue}),
in which the presence of the particle is not considered.
Then, the velocity field without considering the continuity constraint ($\gls{vu}^*$)
is obtained (from a Helmholtz solve) by making use of the forcing \gls{vfibm} to model the presence of the particle.
After that, a Poisson problem is solved in order to obtain the pseudo-pressure $\phi$, with which
the velocity field $\gls{vu}^*$ is corrected to obtain a divergence-free velocity field $\gls{vu}$
and, which serves to update the pressure field.
Finally, the linear and angular position and velocity of the particle are updated.
The complete algorithm reads  

\def\largespaceA{\quad\quad\quad\quad\quad\quad\quad\quad\quad\quad\quad\quad\quad\quad\quad\quad}
\def\largespaceB{\quad\quad\quad\quad\quad\quad\quad\quad\quad\quad\quad\quad\quad\quad\quad\quad}
\makebox[0.8\textwidth][c]{
\begin{minipage}{\textwidth}
\begin{subequations}\label{equ-hybrid-discr-algo2}
\begin{eqnarray} 
%
\notag 
\gls{vue}&=&\gls{vu}^{k-1} +\gls{dt} \left(
                              2\alpha_k\nu\nabla^2\gls{vu_kk}
                             -2\alpha_k\nabla \gls{p_kk}
                             -\gamma_k\left((\gls{vu}\cdot\nabla)\gls{vu}\right)^{k\shortminus 1}\right.
   \\\label{eq:app/ibm/ue}
 & & \left. \,\,\largespaceA -\zeta_k\left((\gls{vu}\cdot\nabla)\gls{vu} \right)^{k\shortminus 2}\right) \,,
\\\label{eq:app/ibm/eu2lag}
\gls{vU_b_e_ml} &=&\sum_{ijk}\gls{vue_b}(\gls{x_ijk}) \delta_h \left(\gls{x_ijk}-\gls{vX_kk_ml}\right)\,\gls{dx}^3 
\,, \quad\forall\,l;\,m;\,\beta
\\
\label{equ-hybrid-discr-algo2-lag-force}
\gls{vF_ml} & = & \frac{ \gls{vUd_kk_ml} - \gls{vU_e_ml} }{\gls{dt}} \,, \quad\qquad\qquad\forall\,l;\,m
\\
\label{equ-hybrid-discr-algo2-eul-force}
{f}_\beta^{(\text{ibm}),k}(\mathbf{x}_{ijk}^{(\beta)})&=&
\sum_{m=1}^{N_p} \sum_{l=1}^{N_L}   \gls{vF_b_ml} \delta_h \left(\gls{x_ijk}-\gls{vX_kk_ml}\right)\Delta V_l^{(m)} 
\,, \quad\forall\,m;\,\beta;\,i;\,j;\,k
\\
\label{equ-hybrid-discr-algo2-predict}
\nabla^2\gls{vu}^\ast-\frac{\gls{vu}^\ast}{\alpha_k\nu\Delta t}&=&
-\frac{1}{\nu\alpha_k}\left(\frac{\tilde{\gls{vu}}}{\Delta t}
+\mathbf{f}^{(\text{ibm}),k} 
\right)
+\nabla^2\gls{vu}^{k-1}
\,,
\\\label{equ-hybrid-discr-algo2-poisson}
\nabla^2\phi&=&\frac{\nabla\cdot\gls{vu}^\ast}{2\alpha_k\Delta t}\,,
\\\label{equ-hybrid-discr-algo2-update-vel}
\gls{vu}^{k}&=&\gls{vu}^\ast-2\alpha_k\Delta
t\nabla\phi\,,
\\\label{equ-hybrid-discr-algo2-update-press}
p^{k}&=&p^{k-1}+\phi-\alpha_k\Delta t\,\nu\nabla^2\phi
 \,,
%
  \\\label{eq:ForceIBM}
  \gls{lagF_km} &=& \sum_l \gls{vF_ml}\Delta V_l^{(m)}    \,, \quad\quad\quad \forall m  \,, 
  \\\label{eq:TorqueIBM}
  \gls{lagT_0_km} &=& \gls{R_kkm}\sum_l (\gls{vX_kk_ml}-\gls{vxp_kk_m}) \gls{vF_ml}\Delta V_l^{(m)}    \,, \quad\quad\quad \forall m \,,
  \\\label{equ-particles-newton-2-present-translation-discrete-u-DEM}
  \frac{\gls{vup_k_m}-\gls{vup_kk_m}}{\gls{dt}} & = & 
                      -\frac{\rho_f}{V_p(\rho_p-\rho_f)}
                      {\mathcal F}^{k,\,(m)}
                      +2\alpha_k\mathbf{g} \,, \qquad\forall\,m
    \\\label{equ-particles-newton-2-present-translation-discrete-x-DEM}
   \frac{\gls{vxp_k_m}-\gls{vxp_kk_m}}{\gls{dt}} & = & 
    \alpha_k\left(\gls{vup_k_m} + \gls{vup_kk_m}\right) \,, \qquad\forall\,m
    \\\notag
    \frac{\gls{vomep_0_km}-\gls{vomep_0_kkm}}{\gls{dt}} &=& %
    -\frac{\gls{rhof}}{\gls{rhop}-\gls{rhof}}\gls{rhop}\,\gls{I0inv} \gls{lagT_0_km}
    -\gamma_k \gls{I0inv}\left( \gls{vomep_0_kkm}\times \gls{I0}\gls{vomep_0_kkm} \right) \\\label{equ-particles-newton-2-present-translation-discrete-om-approx-nonspherical}
     & &\,\quad\quad\quad\quad\quad\quad\quad\quad\quad\quad -\zeta_k    \gls{I0inv}\left( \gls{vomep_0_kkkm}\times \gls{I0}\gls{vomep_0_kkkm} \right), 
    \\\label{equ-particles-newton-2-present-translation-discrete-quat}
    \frac{\gls{quats_km}-\gls{quat_kkm}}{\gls{dt}}&=&\gamma_k\frac{1}{2}\gls{QQuat_kkm}\gls{quat_kkm}%
                                              +\zeta_k\frac{1}{2}\gls{QQuat_kkkm}\gls{quat_kkkm}, 
    \\\label{eq:normquat}
    \gls{quat_km} & = & \gls{quats_km}/\left|\gls{quats_km}\right|
    \\\label{eq:desiredvelocity}
    \gls{vUd_k_ml}  & = &  \gls{vup_k_m} + \gls{RT_km} \left( \gls{vomep_0_km}\times(\gls{vX_0_ml} - \gls{vxp_0_m})\right) 
\end{eqnarray}
\end{subequations}
\end{minipage}
}\\
\\
where $m$ is the index of a given particle ($1\leq m\leq N_p$),
$\beta$ denotes a spatial direction ($1\leq\beta\leq3$),
$\mathbf{X}_l^{(m)}$ is the position of a Lagrangian force point with
index $l$ (where $1\leq l\leq N_l$) attached to the $m$th particle,
$\delta_h$ is the discrete delta function of \citet{roma:1999},
$\mathbf{x}_{ijk}^{(\beta)}$ is the position vector of a node of the
staggered Cartesian fluid grid of the velocity component in the
$x_\beta$ direction with index triplet ``$ijk$'',
$\tilde{{U}}_\beta$ is the velocity in the $x_\beta$-direction
interpolated to a Lagrangian position,
$\mathbf{U}^{(d)}(\mathbf{X}_l)$ is the solid body velocity of the
Lagrangian force point,
$\mathbf{F}(\mathbf{X}_l^{(m)})$ is the immersed boundary force at a
Lagrangian force point,
$\Delta V_l^{(m)}$ is the forcing volume associated to the $l$th
Lagrangian forcing point of the $m$th particle,
$\mathbf{x}_p^{k,\,(m)}$ is the centroid position of the $m$th
particle,
${\mathcal F}^{k,\,(m)}$ is the hydrodynamic force computed
from the sum of the immersed boundary contributions of the $m$th
particle,
${\mathcal T}_b^{k,\,(m)}$ is the analogous hydrodynamic torque contribution 
expressed in the body-fixed coordinate system,
\gls{quats}, \gls{quat} and \gls{QQuat} are the quaternion before and after normalization
and the evolution matrix defined in \S~\ref{sec:ibm/method}, \gls{R} is the rotation matrix (defined below), 
and \gls{I0} is the inertia tensor in the body-fixed reference system.
The set of coefficients $\alpha_k$, $\gamma_k$, $\xi_k$ for a
low-storage scheme leading to second-order temporal accuracy has been given
in \cite{rai:1991}. Please refer to the original publication
\cite{uhlmann:2005} for more details on the algorithm.

It is important to note that the both \gls{lagT_0} and \gls{vomep_0} are expressed
in the body-fixed reference system. 
Therefore, proper transformations through the rotation matrix \gls{R} and 
its transpose $\gls{RT}$ must be applied in steps \eqref{eq:TorqueIBM} and 
\eqref{eq:desiredvelocity}.
Following \cite{tewari:2007} (equation 2.48), the rotation matrix to express 
global (or inertial) coordinates in the body-fixed reference frame is given 
from the quaternion $\gls{quat}$ by
\def\aa{ \gls{q1}^2-\gls{q2}^2-\gls{q3}^2+\gls{q4}^2}
\def\bb{-\gls{q1}^2+\gls{q2}^2-\gls{q3}^2+\gls{q4}^2}
\def\cc{-\gls{q1}^2+\gls{q2}^2+\gls{q3}^2-\gls{q4}^2}
\def\ab{2(\gls{q1}\gls{q2}+\gls{q3}\gls{q4})}
\def\ac{2(\gls{q1}\gls{q3}-\gls{q2}\gls{q4})}
\def\ba{2(\gls{q1}\gls{q2}-\gls{q3}\gls{q4})}
\def\bc{2(\gls{q2}\gls{q3}+\gls{q1}\gls{q4})}
\def\ca{2(\gls{q1}\gls{q3}+\gls{q2}\gls{q4})}
\def\cb{2(\gls{q2}\gls{q3}-\gls{q1}\gls{q4})}
\begin{equation}
\gls{R}=\begin{pmatrix}
\aa   &   \ab   &  \ac \\
\ba   &   \bb   &  \bc \\
\ca   &   \cb   &  \cc \\
\end{pmatrix}.
\end{equation}

Note that the same algorithm \eqref{equ-hybrid-discr-algo2} applies to the case when
we distribute Lagrangian force points throughout the volume occupied by the solid particles. 
The distribution of force points itself in both cases (surface forcing and volume forcing)
is described in \ref{app:lagmesh}.
Let us also note that we have performed extensive tests of the stability and convergence 
of the numerical integration scheme chosen for the rotational rigid body motion, 
including the evaluation of various alternative formulations.
Although more sophisticated approaches with favorable properties exist (e.g.\ based on 
Lie group methods, \cite{sveier:2019}, which circumvent the quaternion 
re-normalization step), the present approach was found to be sufficiently accurate and robust.

\section{Forcing points distribution}
\label{app:lagmesh}

\subsection{Distributing points on the surface of the spheroid}
\label{app:lagmesh/surface}

\definecolor{darkblue}{rgb}{0,0,0.5}

\begin{figure} 
\makebox[\textwidth][c]{ 
\includegraphics[scale=1.0]{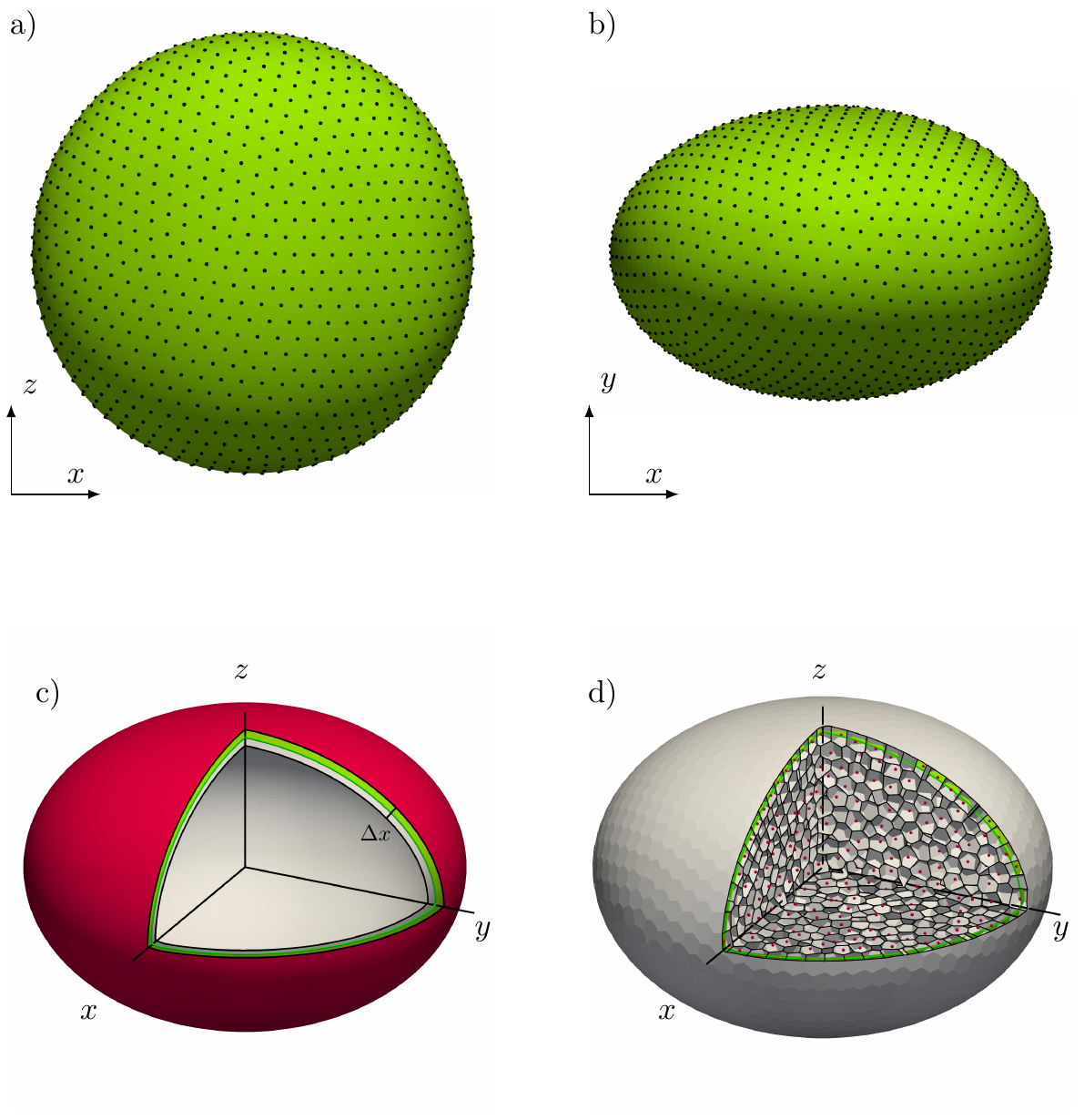} 
}
\caption{a) View along the symmetry axis and b) perpendicular to it of the 
spheroid with aspect ratio $\gls{chi}=1.5$ with the distribution of Lagrangian
markers on its surface.
c) Surface of the spheroid (green) together with the inner (white) and outer (red) 
surfaces that define the shell of thickness \gls{dx}.
d) Lagrangian markers distribution (red points) and the 3D Vorono\"i tessellation
used to distribute the points in the interior of the particle and assign the
Lagrangian markers volume \gls{dv} for the spheroid (green surface).
\label{fig:shellplot}} 
\end{figure}

Here we briefly describe the procedure to distribute the forcing points on the 
surface \review{and interior} of spheroidal particles.
These forcing points are the points in which the explicit estimation of the velocity
is interpolated (see equation \ref{eq:app/ibm/eu2lag}) and  where the forcing
term that models the no-slip boundary condition is computed (equation 
\ref{equ-hybrid-discr-algo2-lag-force}).
For more details the reader is referred to the original description of the 
algorithm \citep{uhlmann:2005}.

In the case of a sphere \cite{uhlmann:2005} defined as an ``even'' distribution 
of points the final state of a simulation in which the total repulsive force 
between all the forcing points, considered as point-particles with a finite and 
equal charge, is minimized (method 2 of their appendix A.2.1).
The repulsive force considered was quadratic with the Euclidean distance between each pair of 
points.
This methodology cannot be directly applied to spheroids because of the lack of 
isotropy in the particle shape.
In the case of oblate spheroids this anisotropy will result in a higher 
concentration of points in the poles of the spheroid.
As expected, the nonuniformity of the points distribution increases as the 
aspect ratio is increased with respect to unity.
In this work we use a similar approach to the one proposed by 
\cite{uhlmann:2005}, but modifying the definition of the repulsive force.
Here, we define the modulus of the repulsive force between two points to be 
proportional to the $N$-th power of the geodesic distance, where $N$ is adjusted
to minimize the nonuniformty of the final distribution.
The direction of the repulsive force between two points is given at each point
by the tangent to the geodesic line that connects both points.
We found that using exponents $N=2$ and $3$ for the definition of the modulus of
the repulsive force results in ``even'' distributions for the spheroids with 
aspect ratio $\gls{chi}=1.1$ and $1.5$, respectively.
In the same fashion as in \cite{uhlmann:2005}, several runs with different 
initial conditions are computed in order to minimize the possibility of finding 
local minima.
Figures \ref{fig:shellplot}a and b show the distribution of $1450$ points on the 
surface of the spheroid with aspect ratio $\gls{chi}=1.5$, which corresponds to 
the Lagrangian markers distribution used for the simultaions with spatial 
resolution $\gls{dd}/\gls{dx}=24$.
The number of points for a given spheroid at a given resolution is obtained as
\begin{equation}
N \approx \frac{V_{shell}}{\gls{dx}^3},
\end{equation}
where $V_{shell}$ represents the volume contained between the inner and outer
surfaces, which are defined as the surfaces separated by a distance of $-\gls{dx}/2$
and $+\gls{dx}/2$ (in the normal direction) from the surface of the spheroid (see figure \ref{fig:shellplot}c).
For small aspect ratios the inner and outer surfaces in figure \ref{fig:shellplot}c can
be approximated by two spheroids of diameter and symmetry axis length of
 ($\gls{dd}-\gls{dx},\gls{aa}-\gls{dx}$) and
 ($\gls{dd}+\gls{dx},\gls{aa}+\gls{dx}$), respectively, resulting in  
\begin{equation}\label{eq:spheroids/dv}
V_{shell}=\frac{\pi \,\gls{dd}^3\,\left(\gls{chi}+\gls{chi}r^2+2\,r^2\right)}{3\,r^3\,\gls{chi}}, 
\end{equation}
where $r=\gls{dd}/\gls{dx}$ is the spatial resolution.
In order to assign the Lagrangian volume $\gls{dv}$ to each forcing point, we generate
a Vorono\"i tesselation on the surface of the spheroid based, again, on geodesic
distances%
\footnote{%
Note that the resultant edges of each cell are nor a straight line, nor a geodesic line,
therefore they are generated numerically to be equidistant (under a threshold and in 
geodesic context) between the two adjacent points. 
}.
The result is that the surface of the spheroid is divided into in cells of fairly even 
distributed area.
The ratio between the area of each cell and the total area of the surface of the
spheroid is used as a factor to assign the volume $\gls{dv}$ from the volume of the 
shell $V_{shell}$.

\subsection{Distributing points throughout the volume of the spheroid}
\label{app:lagmesh/volume}

To distribute points inside the spheroid we use the distribution of points on the surface
of the spheroid from the above mentioned procedure \review{}{(appendix~B.1)},     
and consider these points as fixed.
Then, we randomly fill the interior of the inner surface with $N_i$ points
\begin{equation}
N_i \approx \frac{V_{in}}{\gls{dx}^3},
\end{equation}
where $V_{in}$ is the volume enclosed by the inner surface.
The position of the $N_i$ interior points is iteratively updated following 
Lloyd's method to obtain a centroidal Vorono\"i tesselation
\citep{lloyd:1982,liu:2009}.
\review{}{%
Figure \ref{fig:shellplot}d shows the distribution of points in the volume 
occupied by a spheroid of $\gls{chi}=1.5$ using a spatial resolution of 
$\gls{dd}/\gls{dx}=24$.
}%
This procedure is very attractive because of its simplicity and it can be 
summarized in the following steps:
\begin{enumerate}
   \item Generate a 3D Vorono\"i tessellation of all the forcing points.
   \item Clip the Vorono\"i cells of the points located at the surface of the spheroid
         with the outer surface.
   \item Move each interior point to the centroid of its associated Vorono\"i cell.
   \item Repeat the procedure from step $1$ until some convergence criteria is met. 
         In practice we use the maximum displacement in the most recent step which is required to be below $10^{-8}\gls{dd}$.
   \item Set the Lagrangian volume \gls{dv} equal to the volume of its  Vorono\"i cell.
\end{enumerate}

The implementation to distribute the points on the surface of the spheroid is
heavily based on the library \verb!GeodesicLib! \citep{karney:2013}.
Similarly, the 3D Vorono\"i tesselations to carry out Lloyd's method to 
distribute points in the interior of the spheroid uses the library \verb!voro++!
\citep{rycroft:2009} with a custom implementation of spheroidal walls.

\bibliography{biblio}

\end{document}